\documentclass[12pt]{article}
\usepackage[dvips]{color}
\usepackage{epsfig}
\usepackage{graphicx}
\usepackage{amsmath}

\begin{document}
\begin{center}
\Large{\bf The temperature and entropy corrections on the charged hairy black holes  }\\
\small \vspace{1cm} {{\bf M. Rostami}
\footnote{M.rostami@iau-tnb.ac.ir}},\quad{\bf J. Sadeghi
\footnote{pouriya@ipm.ir}}, \quad {\bf S. Miraboutalebi
\footnote{S-mirabotalebi@iau-tnb.ac.ir}}, \quad {\bf B. Pourhassan
\footnote{b.pourhassan@du.ac.ir}},\\
\vspace{0.5cm}$^{1}$, $^{3}$,   {\it Department of Physics, Tehran Northem Branch,\\
Islamic Azad University, Tehran, Iran}\\
\vspace{0.5cm}$^{2}${\it Department of Physics, Faculty of Basic Sciences,\\
University of Mazandaran, P. O. Box 47416-95447, Babolsar, Iran}\\
\vspace{0.5cm}$^{4}${\it School of Physics, Damghan University, Damghan, 3671641167, Iran}\\
\end{center}\vspace{1.2cm}

\begin{abstract}
In this paper, we consider the first order correction of the entropy and temperature in a charged black hole with a scalar field. Here, we apply such correction for  the different cases of black holes. These corrections are due to the thermal fluctuations of statistical physics.  Also, we take advantage of such corrections and study the $ P - V $ critically and phase transition. Also, we investigate the effect of correction on the critical point and stability of the system. We obtain modified thermodynamics quantities and find effects of thermal fluctuations in the stability of the black hole. We find that stability of black hole is depend on the thermal fluctuations.
Finally, we compare the results of  the  corrected and uncorrected by thermodynamical quantities.\\\\

\end{abstract}
\newpage
\section{Introduction}
Study about the fundamental forces like gravity is one of the main subjects of the elementary particle physics, which help us to the understanding of nature and hence physics laws. As we know, the gravity in (2 + 1)-dimensional space-time is a very important topic of theoretical physics which usually considered as a toy model. These studies began in the early 1980 \cite{1,2,3,4}, by the discovery of BTZ \cite{5} and MTZ \cite{6} black holes. It becomes clear that the three- dimension solution is getting more advantage. The charged black hole with a scalar field in (2 + 1) dimensions already studied by the Ref. \cite{7}. In that case, the scalar field couples to gravity,  and it couples to itself with the self-interacting potential too. Then, the similar black hole with a rotational parameter constructed by the Ref. \cite{8} and then developed by the Ref. \cite{9}. In that case, rotating charged hairy black hole in (2+1) dimensions considered to study Klein-Gordon equation \cite{9-1}. Also, some thermodynamical studies of such kind of black hole may be find in the Refs. \cite{10, 11}.
It has been show that the entropy of large black holes is proportional to the horizon area \cite{12, 13}. It is important to find what happened when the black hole size reduced where thermal fluctuations of the statistical physics yield to logarithmic corrections. It is indeed the first order corrections where the canonical ensemble is stable \cite{14}. The general form of the corrections and their dependence on
the horizon area is a universal feature, which appears in almost all approaches to quantum gravity. Main differences of various approaches are in the correction coefficients. Several works, already done to find the effect of the first order correction on the small black hole thermodynamics \cite{NPB}.
The logarithmic corrections to black hole entropy already obtained by counting microstates in non-perturbative quantum gravity \cite{15, 16}, also by using the Cardy formula \cite{17,18,19}. Moreover, there are other methods where the corrected entropy given by \cite{20,21,22,23,24,25},
\begin{equation}\label{1}
S= S_0 -\frac{\alpha}{2} \ln{S_{0}T_{0}^{2}},
\end{equation}
where $T_{0}$ is Hawking temperature which will be corrected later, and $\alpha $ is a dimensionless parameter, which introduced for the first time by the Ref. \cite{26}, also $ S_0 $ is the Bekenstein-Hawking (BH) entropy. We can trace the effect of  correction using $ \alpha $ and reproduce ordinary thermodynamics when $ \alpha = 0$. There is also other logarithmic corrected entropy as \cite{27},
\begin{equation}\label{2}
S= S_0 -\frac{1}{2} \ln{C_{0} T_{0}^{2}},
\end{equation}
where $C_{0}$ is ordinary specific heat which will also corrected due to the thermal fluctuations.
It is interesting to check that relation (\ref{1}) with $\alpha=1$ and relation (\ref{2}) may yield to similar results for the thermodynamics quantities. It already applied to asymptotically $ AdS $ black holes \cite{28}. In the Ref. \cite{29} modified thermodynamics of a black saturn has been studied by using the corrected entropy (\ref{2}). Then, the corrected thermodynamics of a charged dilatonic black saturn by using both (\ref{1}) and (\ref{2}) investigated by the Ref. \cite{30} and found similar results from both relations.\\
There is also other logarithmic corrected entropy given by,
\begin{equation}\label{S}
S= S_0 -\frac{\alpha}{2} \ln{S_{0}},
\end{equation}
which will be considered in this paper. If we assume $\alpha=3$ then recover the corrected entropy given by \cite{15,20} where rotating BTZ black hole in the Chern-Simons formulation \cite{15} and four dimensional Schwarzschild black hole in loop quantum gravity \cite{20} considered.
In general, it is possible to state that all the different approaches to quantum gravity generate
logarithmic corrections (at the first order approximation) to the area-entropy law of a black hole. It should be note that even though
the leading order corrections to this area-entropy law are logarithmic, the
coefficient of such a term depends on the approach to the quantum gravity. Since
the values of the coefficients depend on the chosen approach to quantum
gravity, we can say that such terms are generated from quantum fluctuations of the space-time
geometry rather than from matter fields on that space-time. Hence, we consider general form given by (\ref{S}) including free parameter $\alpha$ which depends on the given theory. However, it is possible to consider higher order corrections on the black hole entropy \cite{h1,h2,h3,h4,h5,h6}.\\
If the BH entropy corrected, then other thermodynamic quantities also are corrected \cite{31}.
As we know, the thermodynamic stability analyzed by the Refs. \cite{29,30,31,32} under assumption of the fixed temperature. However, it is possible to consider corrected Hawking temperature as \cite{33,34,35,36,37,38},
\begin{equation}\label{T}
T= T_0 (1+\frac{\alpha}{2 S_0}),
\end{equation}
where $T_0$ is ordinary Hawking temperature. We note here the phase transition, and critical point for the black holes have been investigate by several researchers \cite{PV}. The behavior of a black hole compared to van der Waals fluid are studied by the Refs. \cite{39,40,41,42,42-1}. In that case, we use holographic principles and study the  charged  hairy  black hole via a van der Waals fluid.
We use mentioned motivations to investigate logarithmic corrected entropy and temperature  of a  charged hairy  black hole. It is important to note that such thermal fluctuations may be considered as quantum gravity effect \cite{q1,q2,q3}.\\
All the above information gives us motivation to organized paper as follows. In section 2, we review the charged hairy  black holes. In section 3, we consider the effects of quantum corrections on the thermodynamics of charged BTZ  black holes and study the behavior of the  first-order correction on the thermodynamics  of quantities and  stability of black holes. In section 4, we consider an uncharged hairy AdS black hole and derive  the corrected thermodynamics. We also study  the global and local  stability of the uncharged hairy AdS  black hole.
In section 5, we discuss a conformally dressed AdS black hole  and derive  the corrected thermodynamics due to the thermal fluctuations. We also study  the global and local  stability of the conformally  dressed AdS black hole. Finally, we summarize our results by concluding remarks in the last section.

\section{Charged hairy black holes in (2+1) dimensions}
In this section, we consider the solution of Einstein-Maxwell theory which is coupled minimally to scalar field in (2 + 1) dimensions. The hairy black hole is the same solution, and there are lots of texts about this corresponding theory \cite{44,45,46,47,48,49,50,51,52,53,54,55}.
On the other hand, the scalar field may be coupled minimally or nonminimally to gravity. Here, also the self-interacting potential plays important role as $V (\varphi)$ to such model. The above mentioned coupled scalar field lead us to write the corresponding action as,
\begin{equation}
S = \int{d^{3} x \sqrt{-g}\left[ \frac{R}{2} - \frac{1}{2} g^{\alpha \beta} \nabla _{\alpha \varphi} \nabla _{\beta \varphi} - \frac{1}{2} \varepsilon R \varphi^{2} - V(\varphi) - \frac{1}{8} F_{\alpha \beta} F^{\alpha \beta}  \right]},
\end{equation}
where $ \varepsilon = \frac{1}{8}, $ is a constant, and shown the coupling power between gravity and the scalar field. The metric function is given by following expression,
\begin{equation}\label{5}
f(r) = \frac{r^{2}}{\ell^{2}} + 3 \beta -\frac{ Q^{2} }{ 4} + (2 \beta - \frac{ Q^{2} }{ 9}) \frac{B}{r} - Q^2 (\frac{1}{2} + \frac{B}{3 r}) \ln(r),
\end{equation}
where $Q$ is the electric charge, and $ \beta$ is a relation between the black hole charge and mass,
\begin{equation}
\beta = \frac{1}{3} (\frac{Q^2}{4} - M).
\end{equation}

\begin{figure}[h!]
 \begin{center}$
 \begin{array}{cccc}
\includegraphics[width=60 mm]{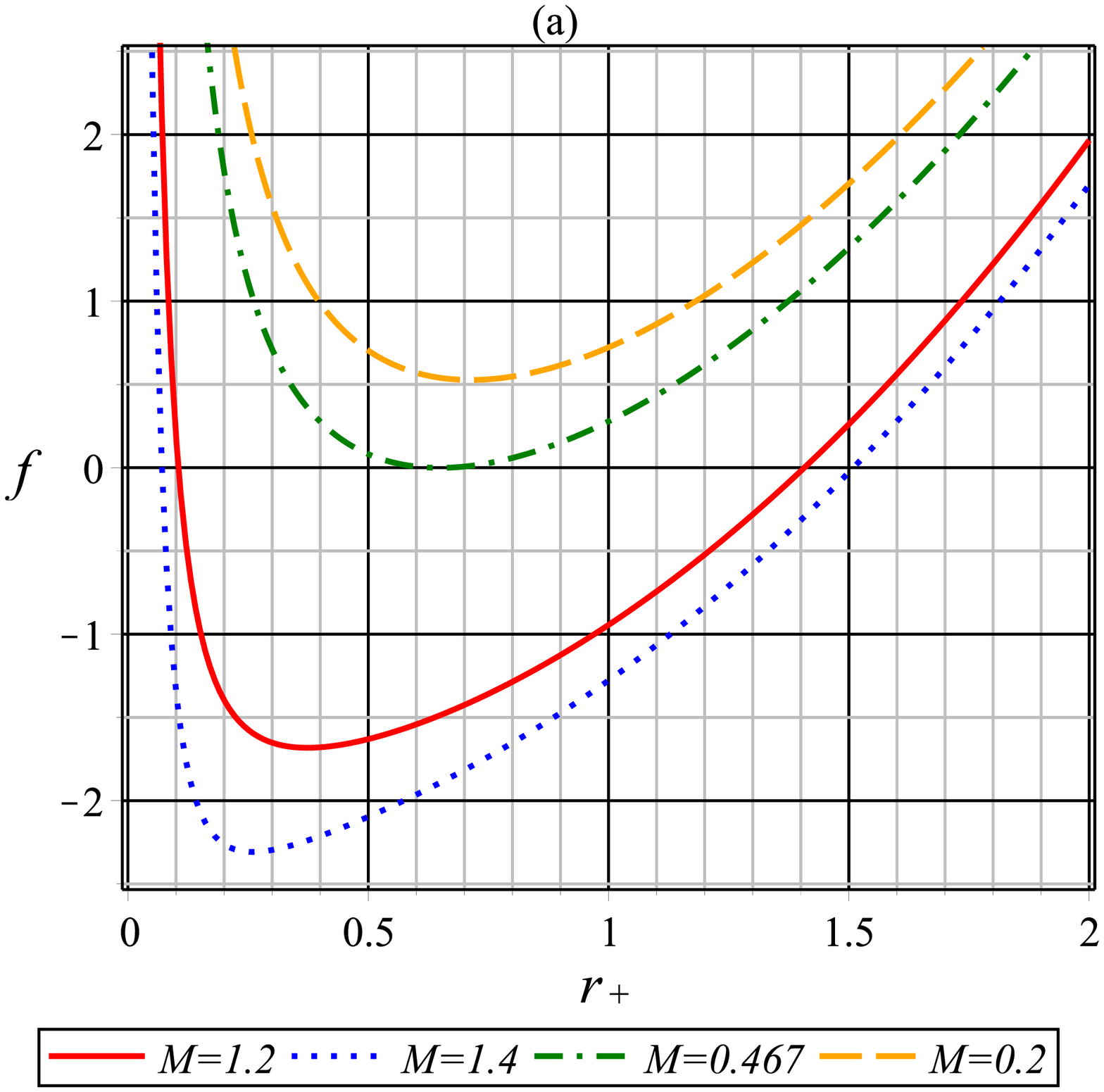}\includegraphics[width=60 mm]{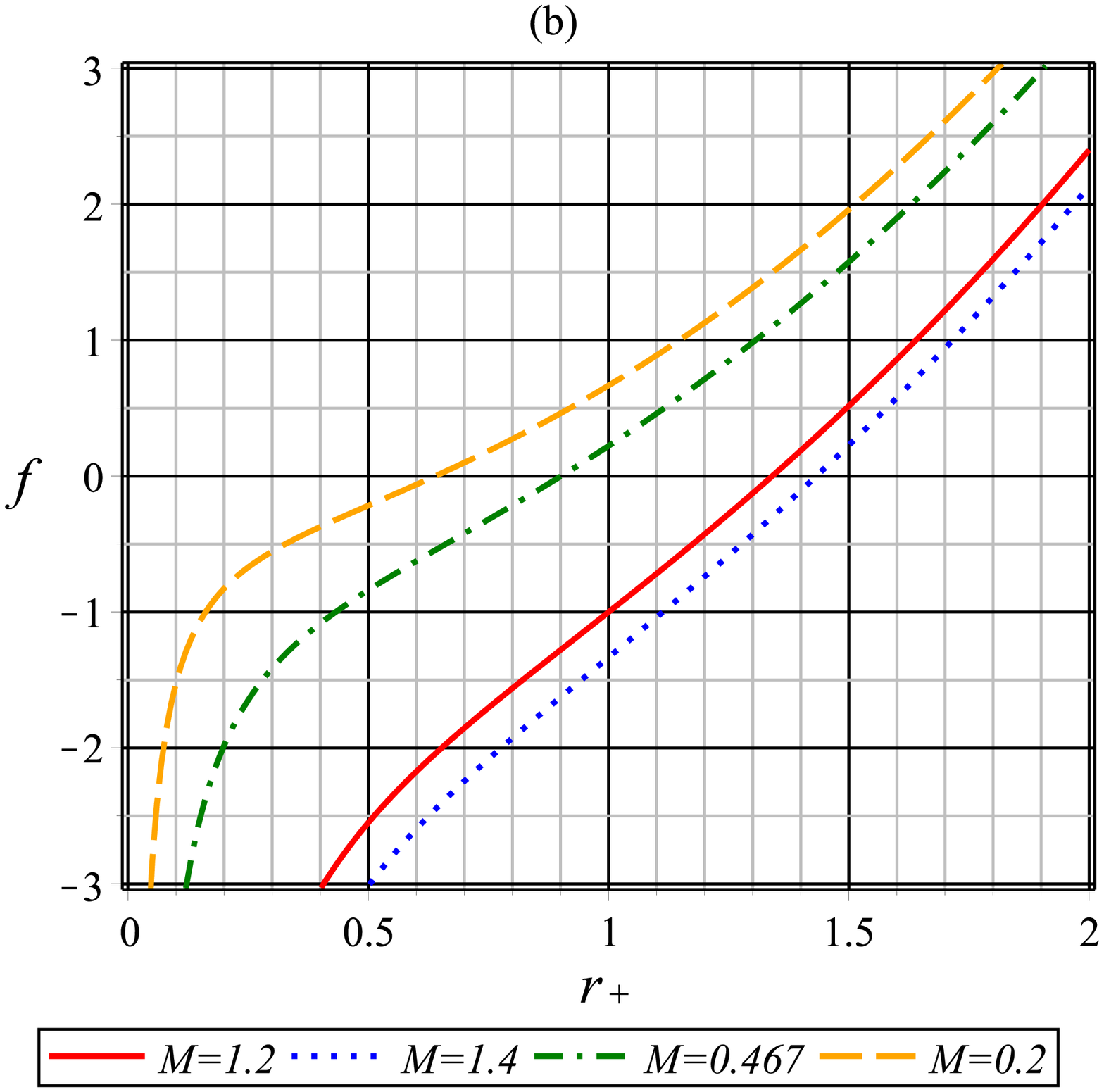}\\
\includegraphics[width=60 mm]{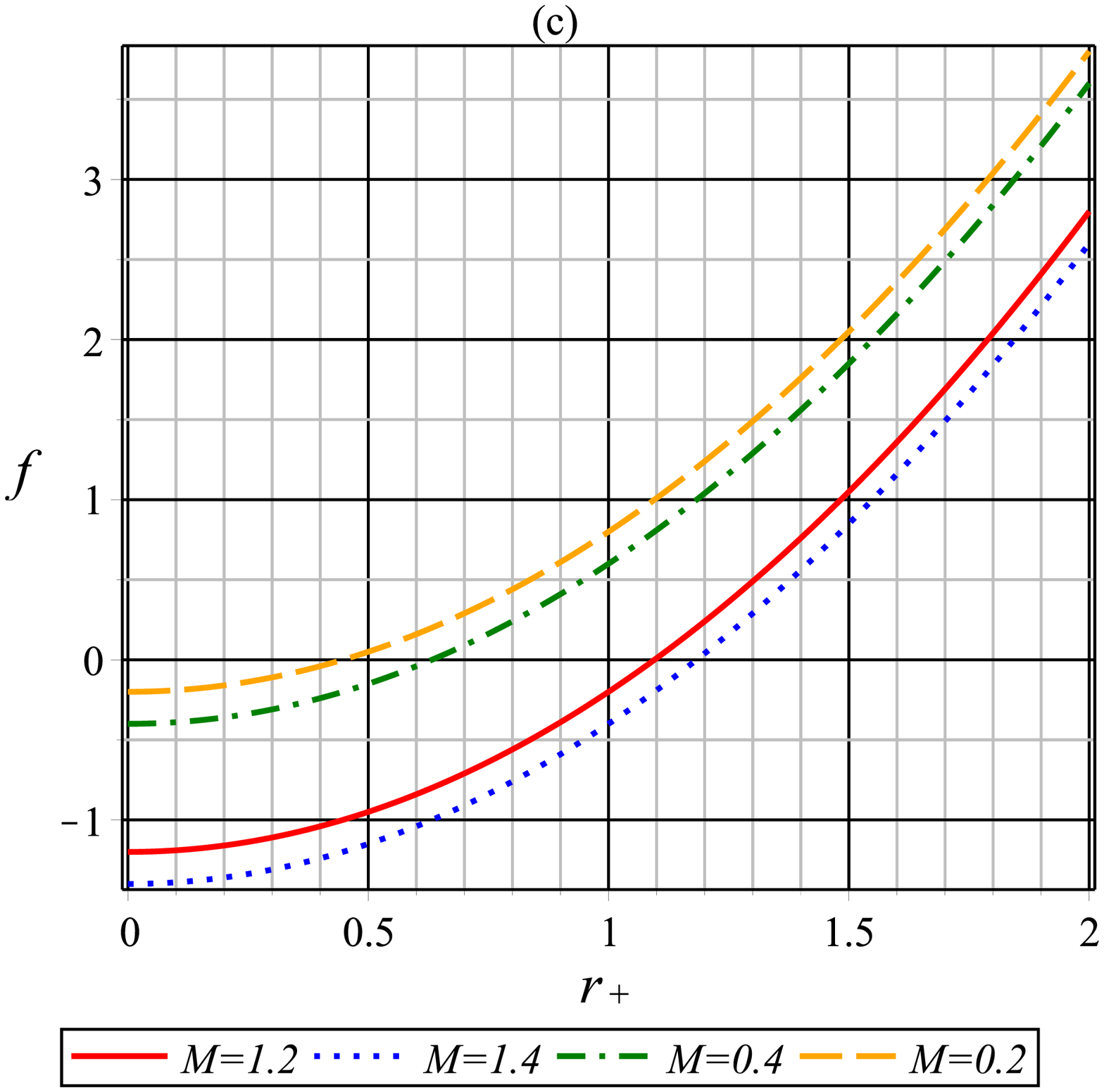}\includegraphics[width=60 mm]{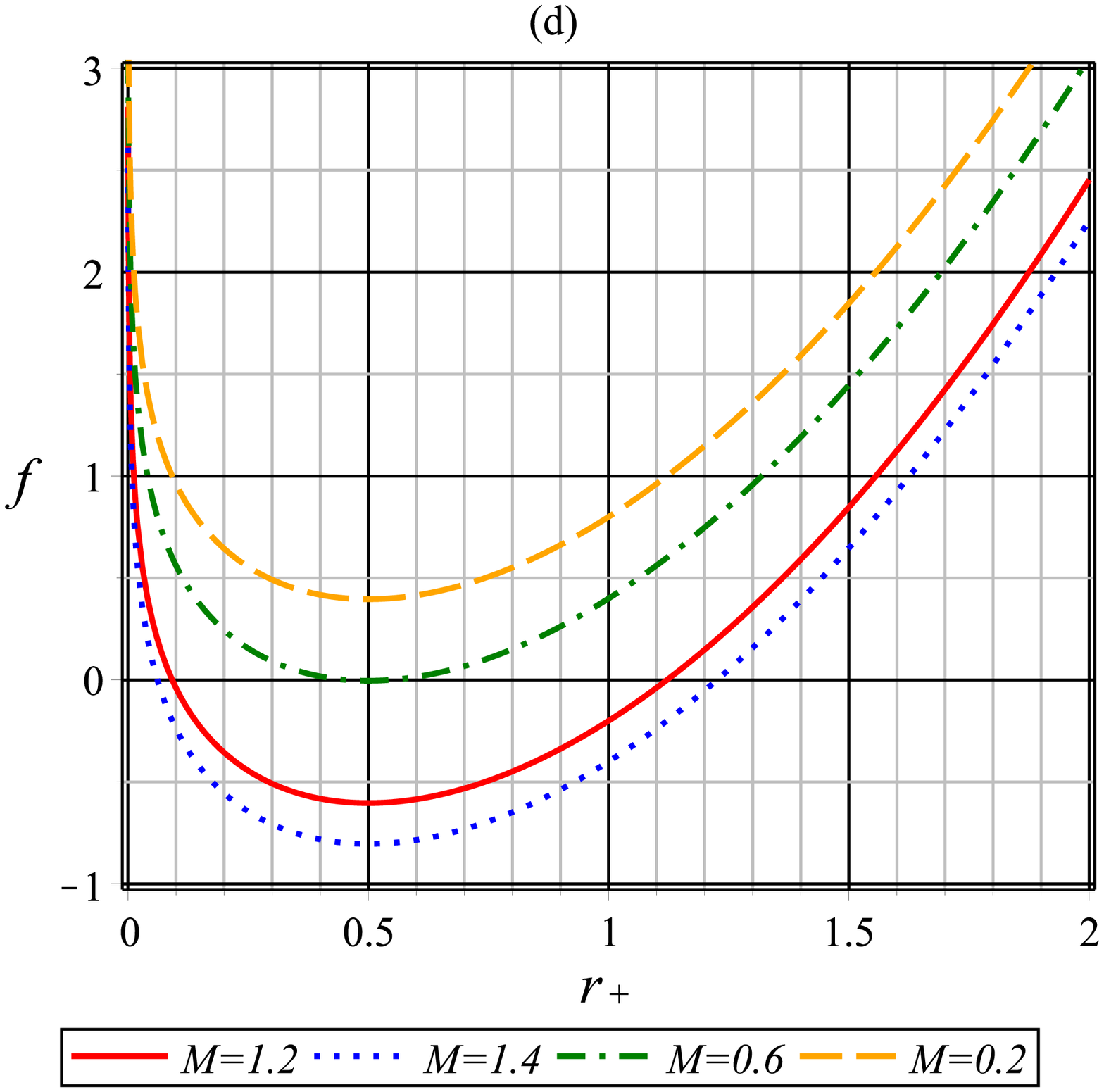}
 \end{array}$
 \end{center}
\caption{Horizon structure of charged hairy black holes in (2+1) dimensions for $\ell=1$. (a) $Q=B=1$ (b) $Q=0$, $B=1$ (c) $Q=B=0$ (d) $Q=1$, $B=0$.}
 \label{fig:1}
\end{figure}

\begin{figure}[h!]
 \begin{center}$
 \begin{array}{cccc}
\includegraphics[width=70 mm]{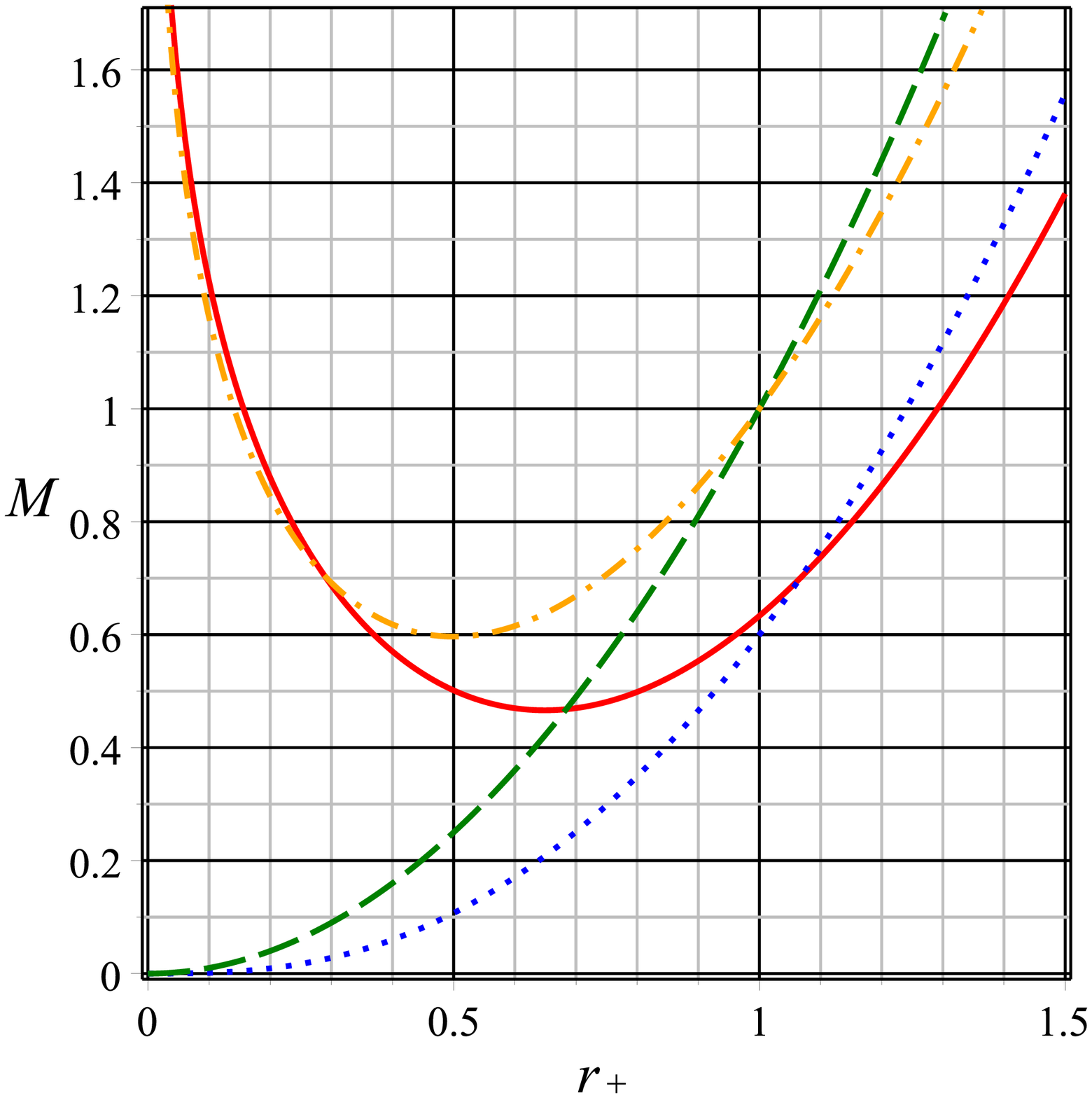}
 \end{array}$
 \end{center}
\caption{Typical behavior of the black hole mass for $\ell=1$. (solid red) $Q=B=1$ (dotted blue) $Q=0$, $B=1$ (dash dotted orange) $Q=B=0$ (dashed green) $Q=1$, $B=0$.}
 \label{fig:2}
\end{figure}

The constant $ \ell $ related to the cosmological constant by $ \Lambda = - \frac{1}{\ell^2}. $  It is negative because smooth black hole horizons can be only in the presence of a negative cosmological constant in (2+1) dimensions, and $r$ explains the radial coordinate. The relation between $B$ and scalar field is as follow,
\begin{equation}
\varphi (r) = \pm \sqrt{\frac{8 B}{r + B}}.
\end{equation}
Graphically, we can see that there are two horizons depend on values of $Q$, $M$, and $B$. In the Fig. \ref{fig:1} (a) we represent typical behavior of $Q=B=\ell=1$ and see that extremal solution given by $M=0.467$. For the cases of $M<0.467$ there is no event horizon and we have only a bare singularity. In the case of $M=1.2$ the event horizon obtained as $r_{+}=1.4$. In the Figs. \ref{fig:1} (b) and (c) we can see that uncharged black hole has only one event horizon.\\
The black hole mass given by,
\begin{equation}\label{mass}
M =\frac{18r_{+}^{3}+Bl^{2}Q^{2}-3(3r_{+}+2B)l^{2}Q^{2}\ln(r_{+})}{6l^{2}(3r_{+}+2B)}.
\end{equation}
In the Fig. \ref{fig:2} we can see behavior of the black hole mass with the event horizon corresponding to four situations of the Fig. \ref{fig:1}.

We will use relation between the black hole mass and event horizon in the thermodynamics study.
These help us to study the modified thermodynamics due to the first order correction of the black hole entropy and temperature.

\section{Corrected thermodynamics for charged BTZ black hole}
If  we assume $ B = 0$ in the equation (\ref{5}),  we will have a black hole solution without a scalar field, so the equation (\ref{5}) reduced to
\begin{equation}\label{8}
f(r) = \frac{r^{2}}{\ell^{2}} - M  -\frac{ Q^{2} }{ 2}  \ln(r).
\end{equation}
It is indeed corresponding to the charged BTZ black hole \cite{BTZ1,BTZ2,BTZ3}. In the Fig. \ref{fig:1} (d) we can see horizon structure of this case. If we choose $Q=1$ the extremal case is corresponding to $M=0.6$. Event horizon of the case $M=1.4$ is about $r_{+}=1.2$.
Therefore, the black hole mass obtained as the follow,
\begin{equation}\label{9}
M =\frac{r^{2}_{+} }{\ell^{2}} -  \frac{ Q^{2} }{ 2}  \ln(r_{+}),
\end{equation}
which is corresponding to dashed green line of the Fig. \ref{fig:2}. Now, by using equations (\ref{8}) and (\ref{9}), the Hawking temperature calculated by,
\begin{equation}
T_{0} =  \frac{1}{4 \pi}\frac{df}{dr} |_{r
= r_{+}} = - \frac{Q^2}{8 \pi r_{+}} + \frac{r_{+}}{2 \pi \ell^2}.
\end{equation}
The charged BTZ black hole entropy is given by,
\begin{equation}\label{12}
S_{0} = 4 \pi r_{+}.
\end{equation}
Here, the negative cosmological constant could interpreted as the positive thermodynamic pressure,
\begin{equation}
P = - \frac{\Lambda}{8 \pi} = \frac{3}{8 \pi \ell^{2}}.
\end{equation}
Utilizing the relations (\ref{S}) and (\ref{T}), the first-order corrected entropy and temperature  for the charged BTZ black hole is obtained by the following  equations,
\begin{equation}
S= 4 \pi r_{+} - \frac{\alpha}{2} \ln{ \left( 4 \pi r_{+} \right) },
\end{equation}
and,
\begin{equation}
T= - \frac{Q^2}{8 \pi r_{+}} +  \frac{r_{+}}{2 \pi \ell^2} - \frac{ \alpha Q^2}{(8 \pi r_{+})^2 } + \frac{\alpha}{(4 \pi  \ell)^{2}}.
\end{equation}

\begin{figure}[h!]
 \begin{center}$
 \begin{array}{cccc}
\includegraphics[width=60 mm]{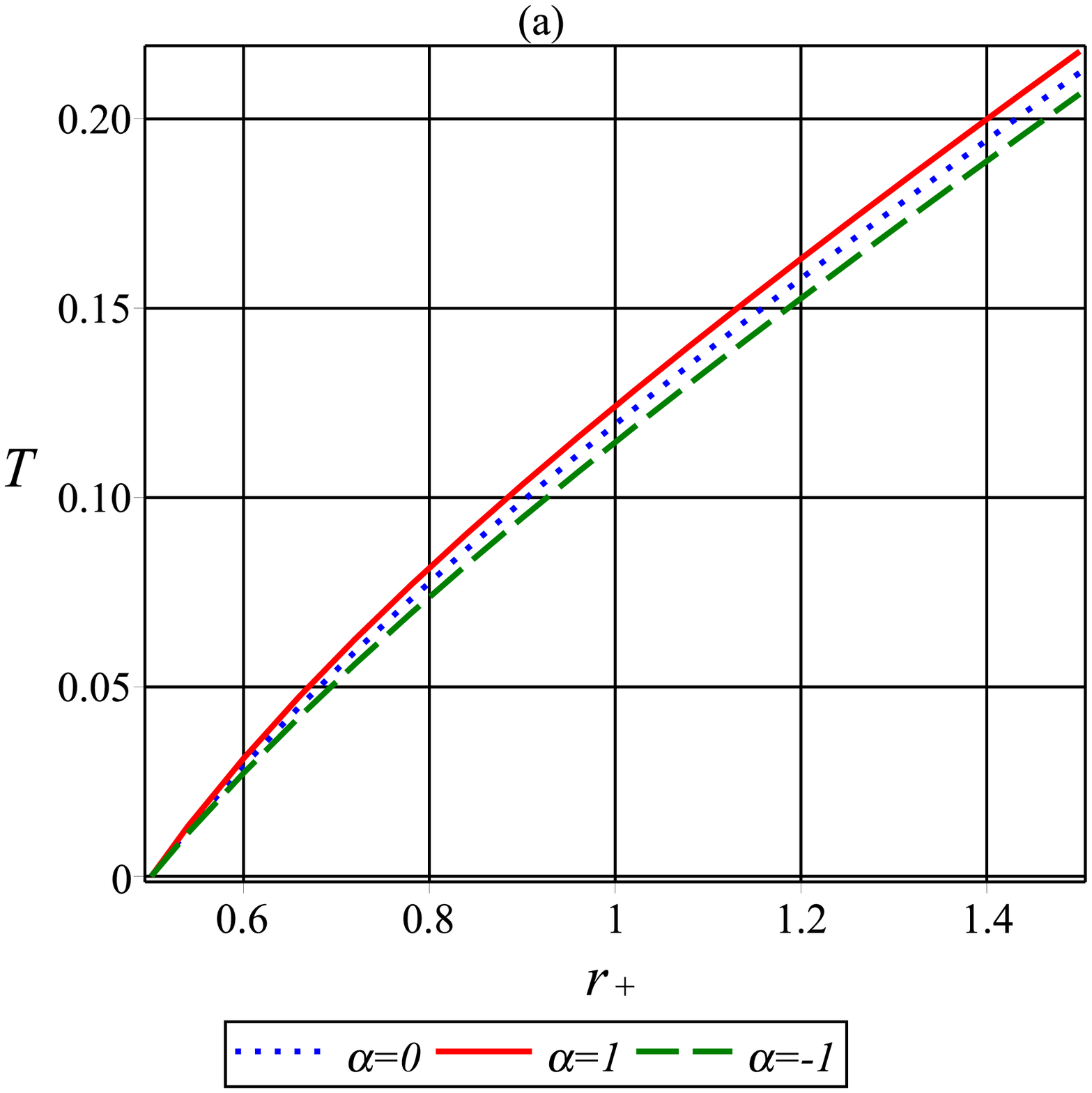}\includegraphics[width=60 mm]{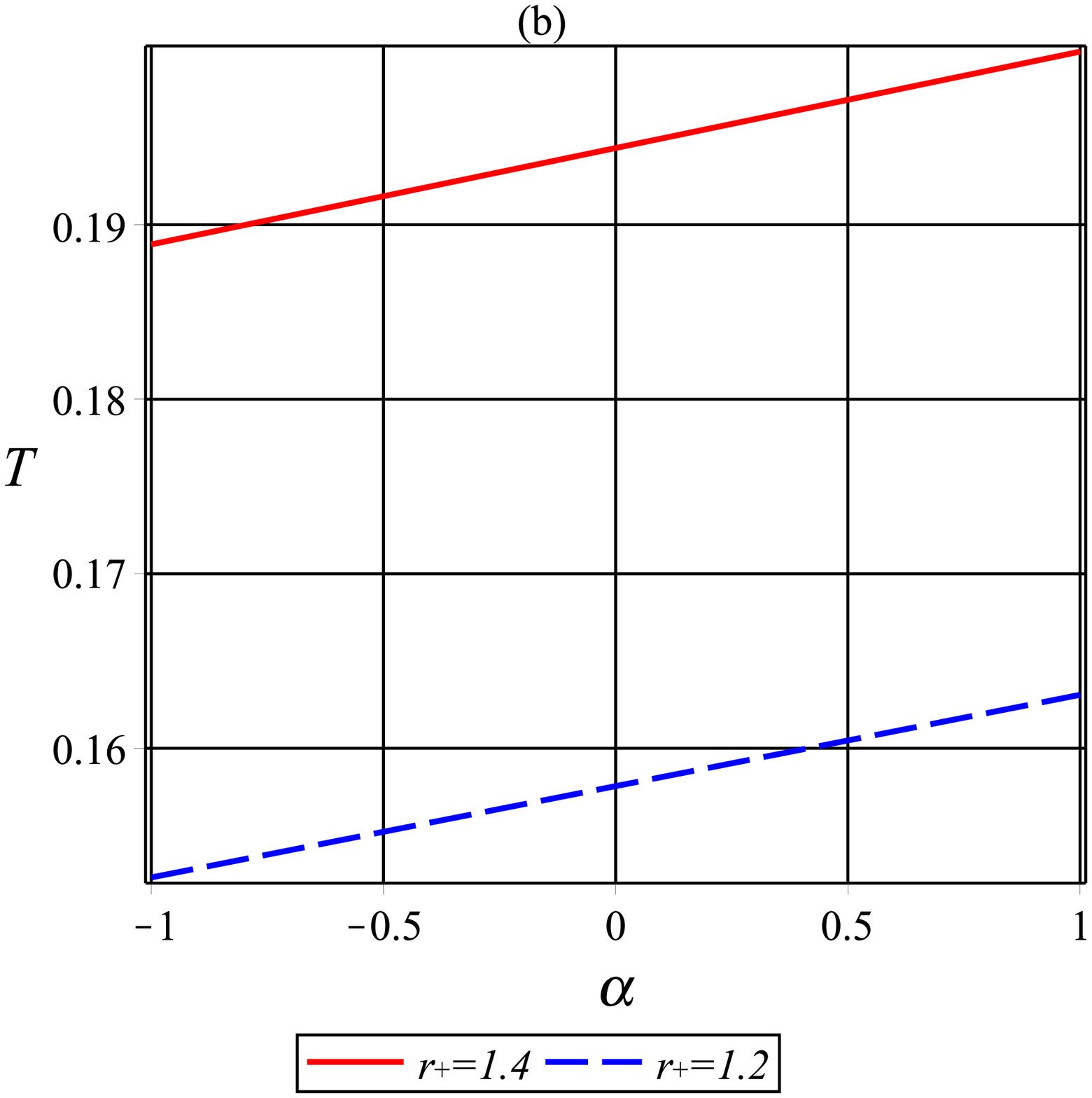}
 \end{array}$
 \end{center}
\caption{Typical behavior of the corrected Hawking temperature of charged BTZ black hole for $Q=\ell=1$. (a) in terms of $r_{+}$ (b) in terms of correction coefficient.}
 \label{fig:3}
\end{figure}

\begin{figure}[h!]
 \begin{center}$
 \begin{array}{cccc}
\includegraphics[width=60 mm]{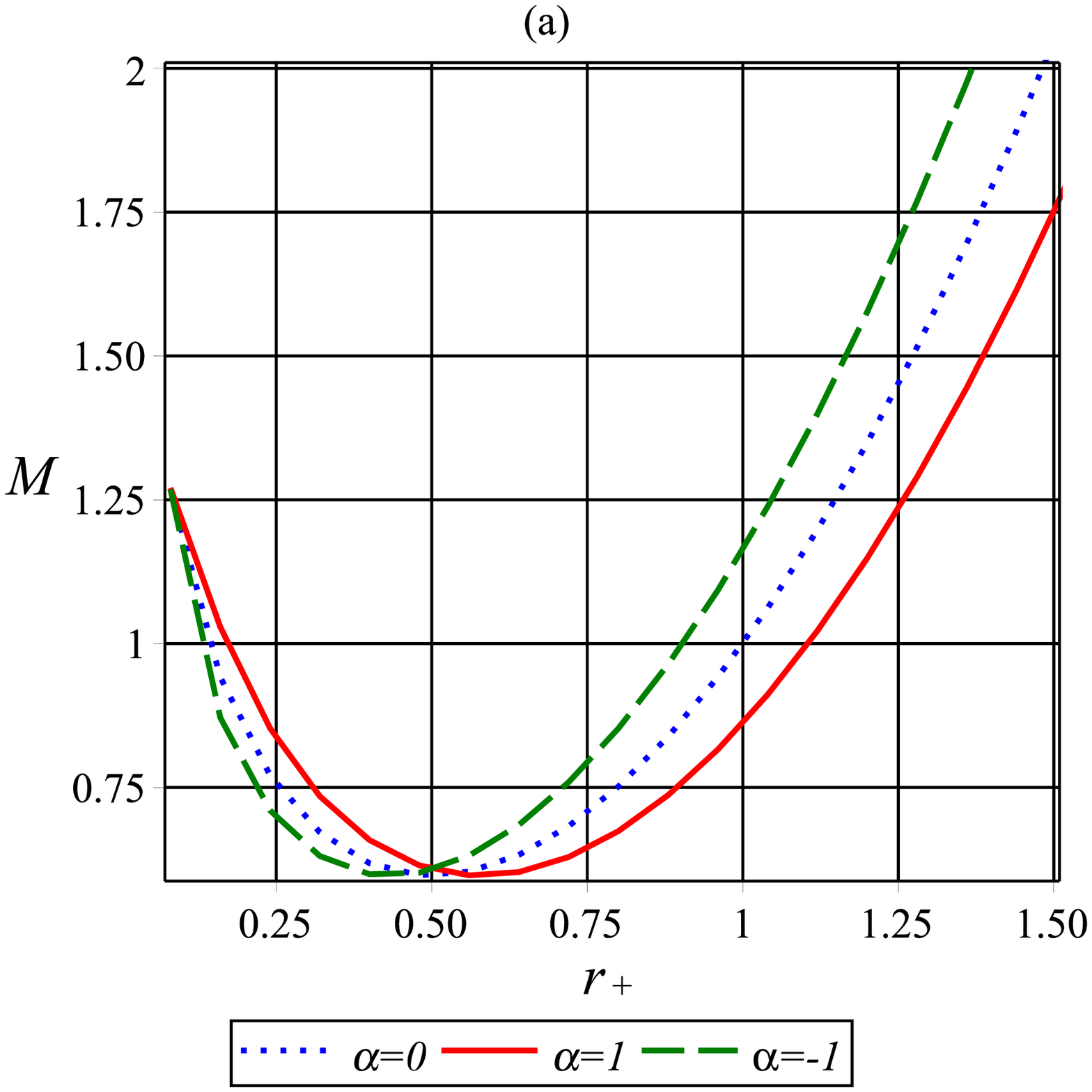}\includegraphics[width=60 mm]{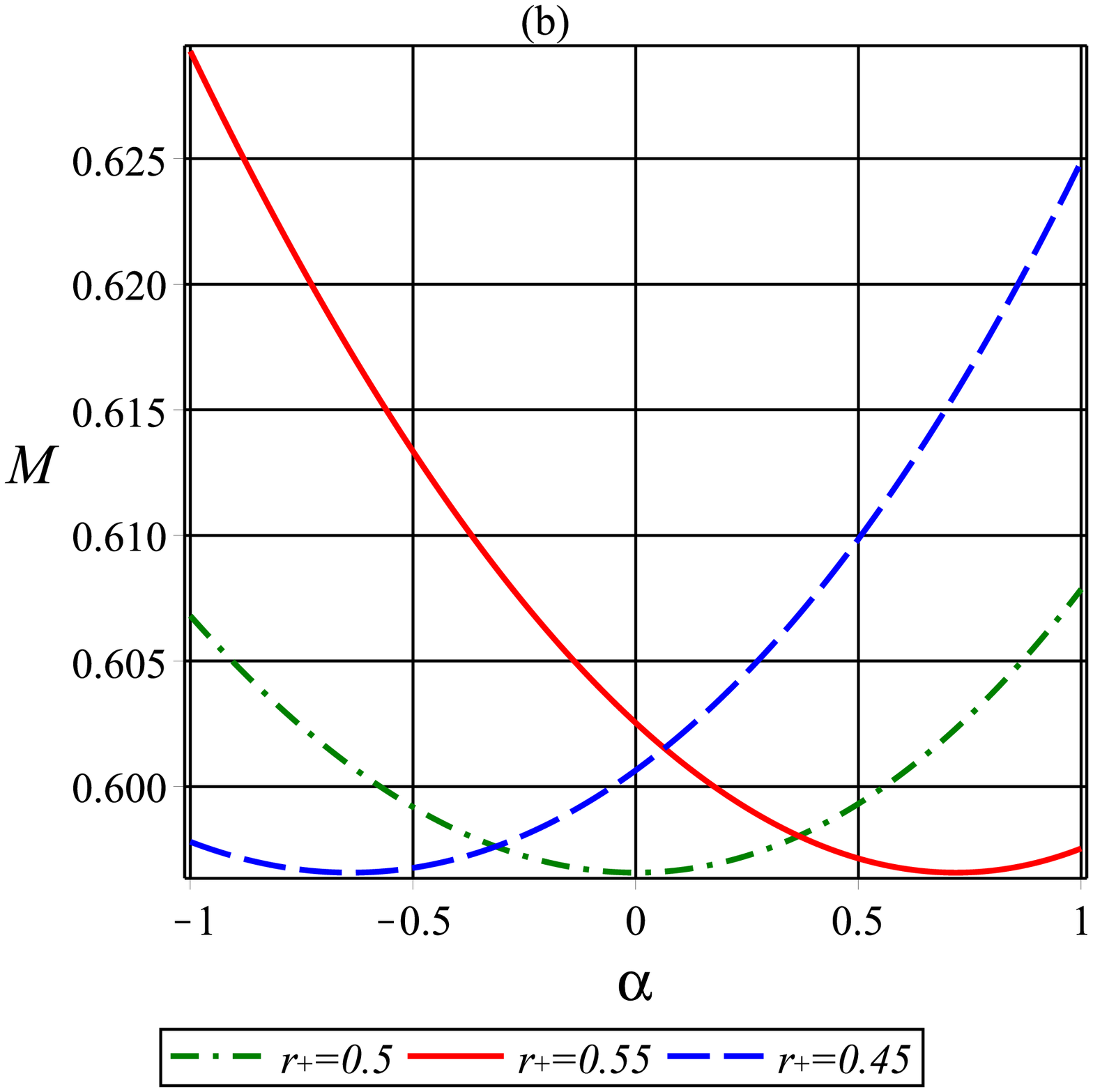}
 \end{array}$
 \end{center}
\caption{Typical behavior of the charged BTZ black hole mass for $Q=\ell=1$. (a) in terms of $r_{+}$ (b) in terms of correction coefficient.}
 \label{fig:4}
\end{figure}

Plots of the Fig. \ref{fig:3} show that the Hawking temperature is increasing function of $\alpha$. It means that thermal fluctuations increase value of the Hawking temperature.\\
From equation (\ref{9}), the corrected physical mass for the charged  BTZ  black hole is obtained as,
\begin{equation}
M = \left(\frac{  8 \pi r_{+} - \alpha \ln (4 \pi r_{+} ) }{ 8 \pi \ell} \right)^{2} + \frac{Q^{2}}{2} \ln{(8 \pi)} -  \frac{Q^{2}}{2} \ln{(8 \pi r_{+} - \alpha \ln (4 \pi r_{+} ) )} ,
\end{equation}
We plot it in Fig. \ref{fig:4} in terms of horizon radius as Fig. \ref{fig:4} (a) and in terms of correction coefficient as Fig. \ref{fig:4} (b). For the selected values ($Q=\ell=1$) we know that $r_{+}\geq0.5$ (see Fig. \ref{fig:1} (d)) hence black hole mass is decreasing function of $\alpha$.\\

The first law of thermodynamic for the black hole reads,
\begin{equation}\label{fl}
dM = T dS + V dP + \phi dQ,
\end{equation}
where $V$ is the thermodynamic volume, and $ \phi $ is the electric potential. In that case, by using entropy and temperature \cite{EPJC} one can obtain Helmholtz free energy as,
\begin{eqnarray}
F = -\int{S dT} =  &-& \frac{Q^2}{2} \ln(r_+) - \frac{r^{2}_{+} }{\ell^{2}}\nonumber\\
&+&\frac{\alpha}{16 \pi^{2} \ell^2 r_{+}}[\pi Q^2 \ell^2 (1-\ln(4 \pi r_{+})) + 4 \pi r^{2}_{+} \ln(4 \pi r_{+})  - r^{2}_{+} ]\nonumber\\
&-& \frac{\alpha^{2} Q^2 }{(16 \pi)^2 r^{2}_{+}}[1 + 2 \ln(4 \pi r_{+})].
\end{eqnarray}
Using definition, $ E =  F + S T  $, the corrected internal energy is calculated as,
\begin{eqnarray}
E &=& \frac{r^{2}_{+} }{\ell^{2}}  - \frac{Q^2}{2}( 1 +  \ln(4 \pi r_+) ) - \frac{\alpha}{16 \pi^2 \ell^2 } (1 - 4 \pi) r_{+}\nonumber\\
&+& \frac{\alpha^{2}  }{(16 \pi \ell)^2 r^{2}_{+}}[\ell^2 Q^2(1 - 2 \ln(4 \pi r_{+})) - 8 r^{2}_{+}  \ln(4 \pi r_{+})]
\end{eqnarray}
Also, one can obtain the enthalpy as following,
\begin{eqnarray}
H &=& E + P V  = \frac{r^{2}_{+} }{\ell^{2}} - \frac{r^{3}_{+} }{2 \ell^{2}}  - \frac{Q^2}{2}( 1 +  \ln(4 \pi r_+) )- \frac{\alpha}{16 \pi^2 \ell^2 } (1 - 4 \pi) r_{+}\nonumber\\
&+& \frac{\alpha^{2}  }{(16 \pi \ell)^2 r^{2}_{+}}[\ell^2 Q^2(1 - 2 \ln(4 \pi r_{+})) - 8 r^{2}_{+}  \ln(4 \pi r_{+})].
\end{eqnarray}
Finally, the quantum corrected Gibbs free energy given by,
\begin{eqnarray}\label{G}
G &=& H - ST =  - \frac{r^{2}_{+} }{\ell^{2}} - \frac{r^{3}_{+} }{2 \ell^{2}}  - \frac{Q^2}{2}  \ln(4 \pi r_+)\nonumber\\
&+& \frac{\alpha}{16 \pi^2 \ell^2  r_{+}} [\pi  \ell^2 Q^2 (1 -  \ln(4 \pi r_{+})) + 4 \pi  r^{2}_{+}  \ln(4 \pi r_{+}) - r^{2} _{+}]\nonumber\\
&-& \frac{\alpha^{2} Q^2 }{(16 \pi r_{+})^2} [1 + 2 \ln(4 \pi r_{+})].
\end{eqnarray}

\begin{figure}[h!]
 \begin{center}$
 \begin{array}{cccc}
\includegraphics[width=50 mm]{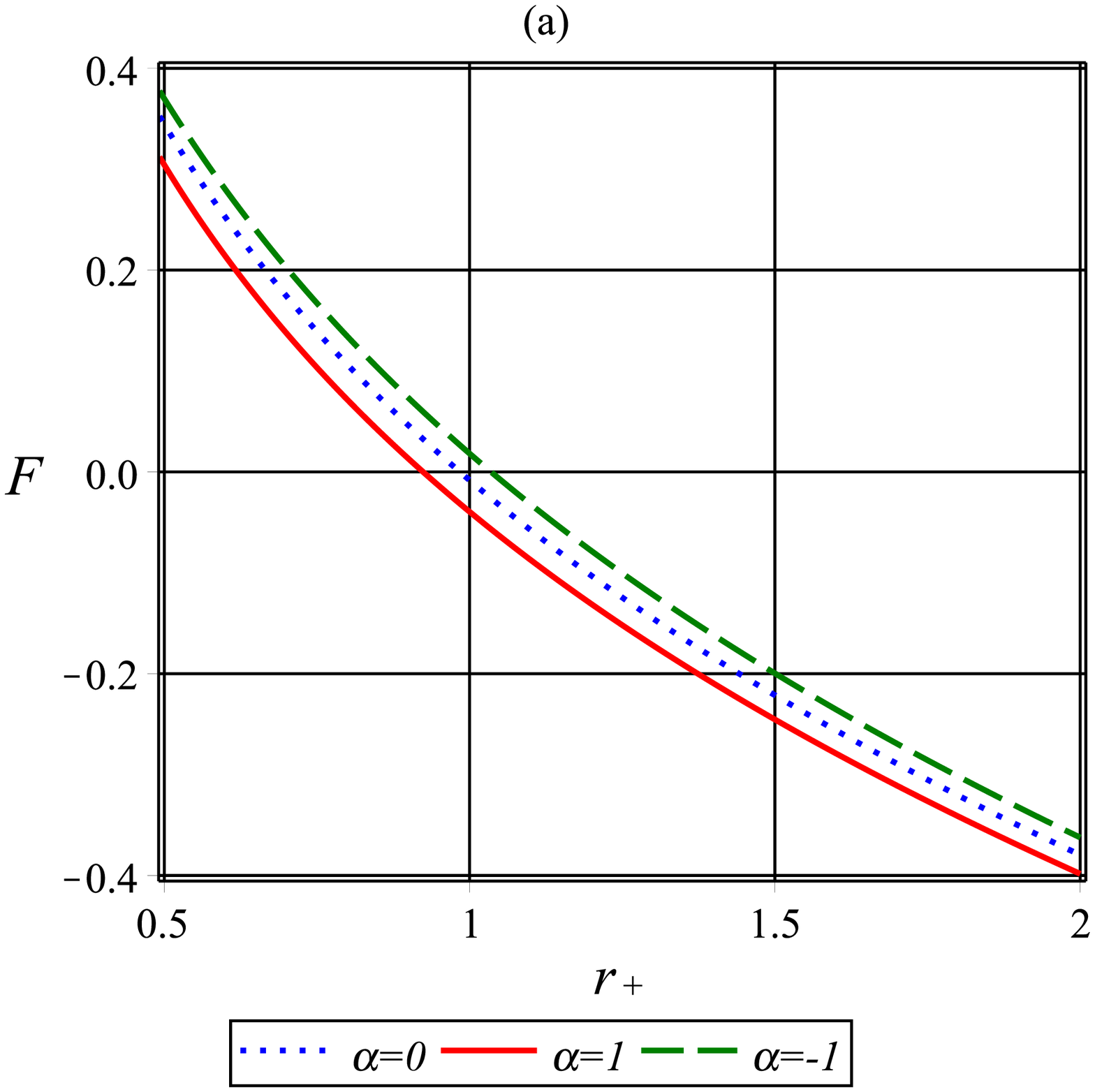}\includegraphics[width=50 mm]{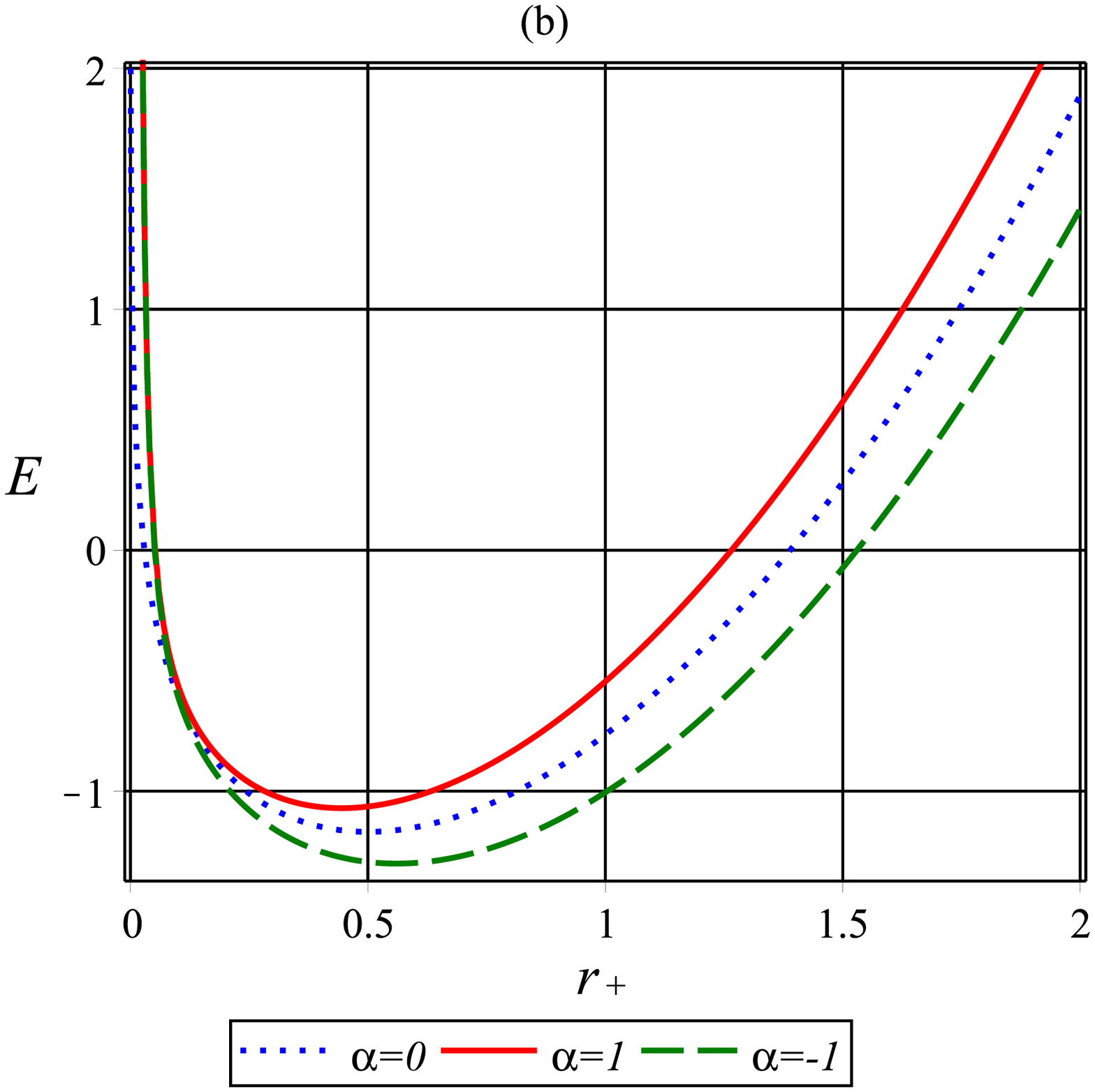}\\
\includegraphics[width=50 mm]{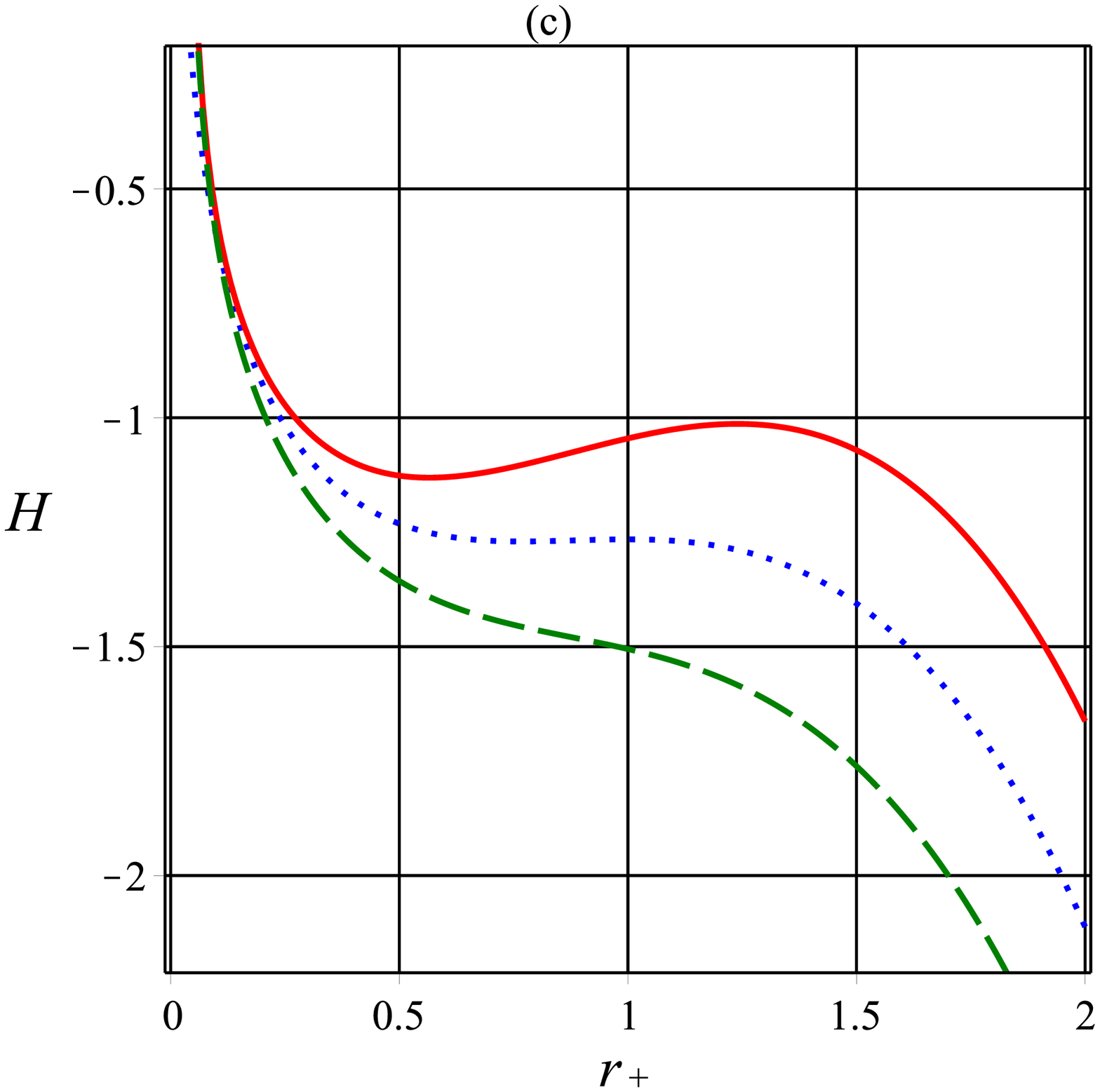}\includegraphics[width=50 mm]{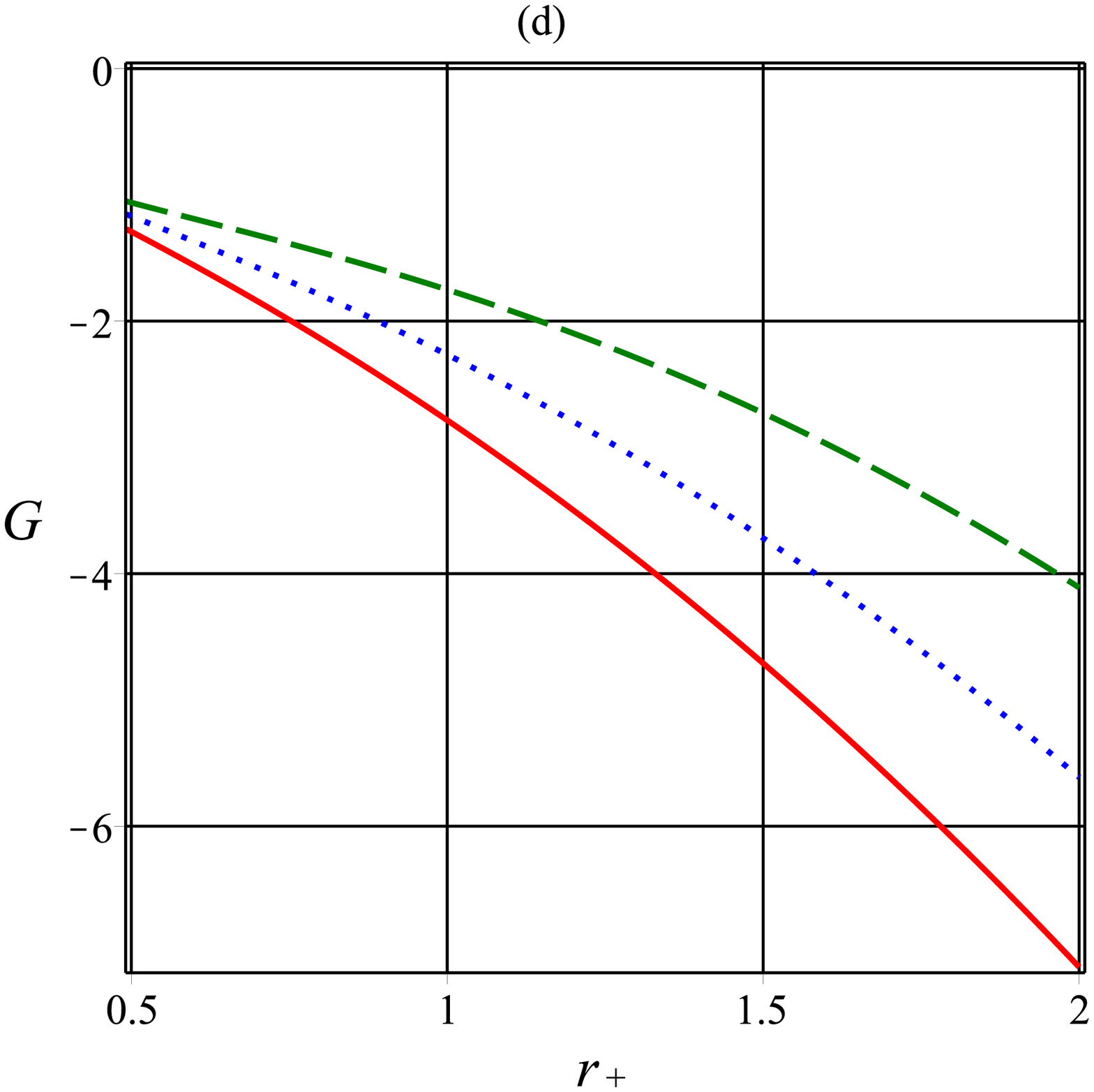}
 \end{array}$
 \end{center}
\caption{Typical behavior of (a) Helmholtz free energy (b) internal energy (c) enthalpy (d) Gibbs free energy in terms of $r_{+}$ for the charged BTZ black hole with $Q=\ell=1$.}
 \label{fig:5}
\end{figure}

In the plots of Fig. \ref{fig:5} we can see typical behavior of Helmholtz free energy, internal energy, enthalpy and Gibbs free energy to find the effect of quantum correction. From Fig. \ref{fig:5} (a) one can find that Helmholtz free energy is decreasing function of correction coefficient. In the Fig. \ref{fig:5} (b) we can see that internal energy is minimum corresponding to the extremal black hole. Hence, for the cases of $r_{+}\geq0.5$ internal energy is increasing function of the correction parameter. In the Fig. \ref{fig:5} (c) it is illustrated that the corrected enthalpy has a maximum (with negative value) and enthalpy is increased due to the thermal fluctuations. Extremum of the enthalpy may sound stability of the system. Finally, Fig. \ref{fig:5} (d) represent variation of Gibbs free energy with the event horizon, and we can see that net value of Gibbs free energy increase with positive $\alpha$.\\
We will discuss the critical points and the stability of the charged BTZ black hole by using Gibbs free energy and specific heat in the next section.

\subsection{The critical points and stability of charged BTZ black bole}
Now, we are going  to consider the stability condition for the corresponding system. In that case, we need two quantities which play an important role in the study of the stability system as Gibbs free energy and heat capacity.
The critical points $(r_{+,c}, P_{c}, T_{c})$ in the phase transition of the charged BTZ black hole with the corrected entropy and temperature obtained by the following conditions,
\begin{equation}\label{Con}
\frac{\partial T}{\partial r_{+}} = \frac{\partial ^{2}T}{\partial r_{+}^{2}} = 0,
\end{equation}
which yields to the following relations,
\begin{equation}
\frac{4 \pi r_{+} \ell^{2} Q^{2} +  \alpha\ell^{2} Q^{2} +16 \pi  r_{+}^{3}}{32 \pi^{2}  \ell^{2} r_{+}^{3}} = 0,
\end{equation}
and,
\begin{equation}
\frac{ 8 \pi r_{+} + 3 \alpha   }{32 \pi^{2}  r_{+}^{4} } = 0.
\end{equation}
Using these conditions one obtain the critical values,
\begin{equation}
r_{+,c} = \frac{-3 \alpha }{8\pi},
\end{equation}
and,
\begin{equation}
P_{c} = \frac{2 \pi Q^{2}}{9  \alpha^{2} }, \qquad  T_{c} = \frac{4 \pi Q^{2}}{27 \alpha^{2}},
\end{equation}
the expression for specific volume is given by,
\begin{equation}
\nu_{c} = 6 \frac{V}{A}  = \frac{3 \alpha }{8\pi}.
\end{equation}
Now, the critical ratio is calculated as,
\begin{equation}
\frac{P_{c} \nu_{c} }{T_{c}} = \frac{3 \alpha}{32 \pi}.
\end{equation}
As we see $ \nu_{c} $ and $ \frac{P_{c} \nu_{c}}{T_{c}} $ were increased but $T_{c}$ and $ P_{c} $  are decreased by increasing $ \alpha $. Also we
observe when $  \alpha$ is negative $ P_{c} $ and $T_{c}$ don't change but  $ \nu_{c} $ and $ \frac{P_{c} \nu_{c}}{T_{c}} $ decrease.
If $ \alpha = 4 \pi $,  the above product will be as a usual relation $ \frac{P_{c} \nu_{c} }{T_{c}} = \frac{3 }{8 }. $ This show when we have charged BTZ black hole the $ \nu_{c} $  changed to form of $ \frac{3}{2} $ without correction.\\
There is another method to study critical behavior in the extended phase space which is based on the Gibbs free energy \cite {57, 58}.
When the Gibbs free energy is negative $ (G < 0)$, the system has global stability. To discuss such global stability of the black hole, we need to calculate the Gibbs free energy  in the presence of quantum  corrected entropy and corrected temperature which is given by the equation (\ref{G}). In the Fig. \ref{fig:5} (d) we can see that Gibbs free energy is totally negative for selected value of the correction parameter hence global stability established even in presence of the logarithmic correction.\\
The specific heat is an important measurable physical quantity and  also is determine the local thermodynamic stability of the system. The mentioned specific heat has the following relation with  corrected entropy $S$ and  temperature $T$,
\begin{equation}
C = T \left(\frac{dS}{dT} \right).
\end{equation}
In the cases of  $C > 0$ ($C < 0$) the black hole is in stable (unstable) phases respectively. So, $C = 0$ with asymptotic behavior corresponds to the phase transition of the van der Waals fluid. By using of the  above equation, we can obtain specific heat as,
\begin{equation}
C = \frac{(\alpha - 8 \pi r_{+} )}{2}\left[ \frac{ \ell^{2}Q^{2}(16 \pi r_{+} + \alpha) - 48 \pi r^{3}_{+} - 4 \alpha r^{2}}{\ell^{2}Q^{2}(8 \pi r_{+} + \alpha) + 64 \pi r^{3}_{+} }  \right] ,
\end{equation}

\begin{figure}[h!]
 \begin{center}$
 \begin{array}{cccc}
\includegraphics[width=70 mm]{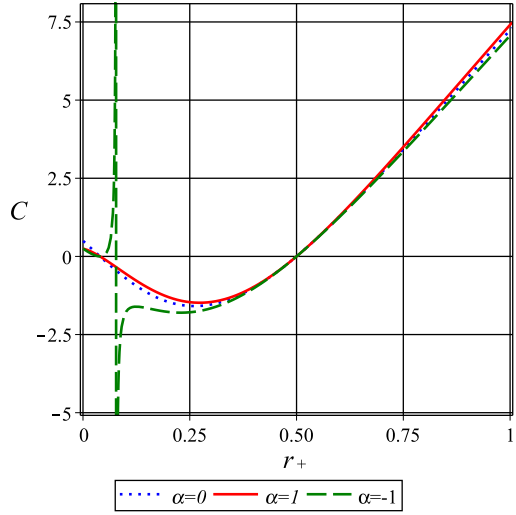}
 \end{array}$
 \end{center}
\caption{Typical behavior of specific heat of the charged BTZ black hole mass with $Q=\ell=1$.}
 \label{fig:6}
\end{figure}

In Fig. \ref{fig:6}, we observe the behavior of the heat capacity for differen values of the correction parameter $ \alpha $. Here, we can see the effects of the logarithmic corrected entropy and corrected temperature on the stability of charged BTZ black hole. We can see that the small positive values $ \alpha $ have no significant effect on the specific heat hence on the stability of the black hole. However, for the negative values of correction coefficient we can see phase transition which is corresponding to the asymptotic behavior (dashed green line of Fig. \ref{fig:6}).

\section{Corrected thermodynamics for uncharged hairy AdS black hole}
Now, if we assume $ Q = 0 $ in equation (\ref{5}), in that case,  the metric function reduced to the following expression,
\begin{equation}
f(r) = - M(1 +\frac{2 B}{3 r} ) + \frac{r^{2}}{\ell^{2}}
\end{equation}
where the physical mass is,
\begin{equation}\label{31}
M =\frac{r^{2}_{+} }{\ell^{2} (1 +\frac{2 B}{3 r_{+}} )} .
\end{equation}
Horizon structure of this case plotted by the Fig. \ref{fig:1} (b).
The thermodynamic volume can be calculated by,
\begin{equation}
V =\left( \frac{\partial M }{\partial P} \right) _{S} = - \frac{ 8 \pi r_{+} ^{3}}{(3 r_{+} + 2 B)},
\end{equation}
So,  the corresponding  temperature will be,
\begin{equation}
T_{0} = \frac{9 S_0^2(S_0 + 4 \pi B )}{8 \pi^2 \ell^2 (3 S_0 + 8 \pi B )^2} ,
\end{equation}
the  entropy density of uncharged hairy black holes is also given by (\ref{12}).
Utilizing the relations (\ref{S}) and (\ref{T}), the first-order corrected entropy and corrected temperature  for the uncharged hairy AdS black hole is computed as,
\begin{equation}
S= 4 \pi r_{+} - \frac{\alpha}{2} \ln{4 \pi r_{+}},
\end{equation}
and,
\begin{equation}
T= \frac{ 9 S_0 (2 S_{0} + \alpha )(S_0 + 4 \pi B )}{16 \pi^2 \ell^2 (3 S_0 + 8 \pi B )^2}.
\end{equation}

\begin{figure}[h!]
 \begin{center}$
 \begin{array}{cccc}
\includegraphics[width=60 mm]{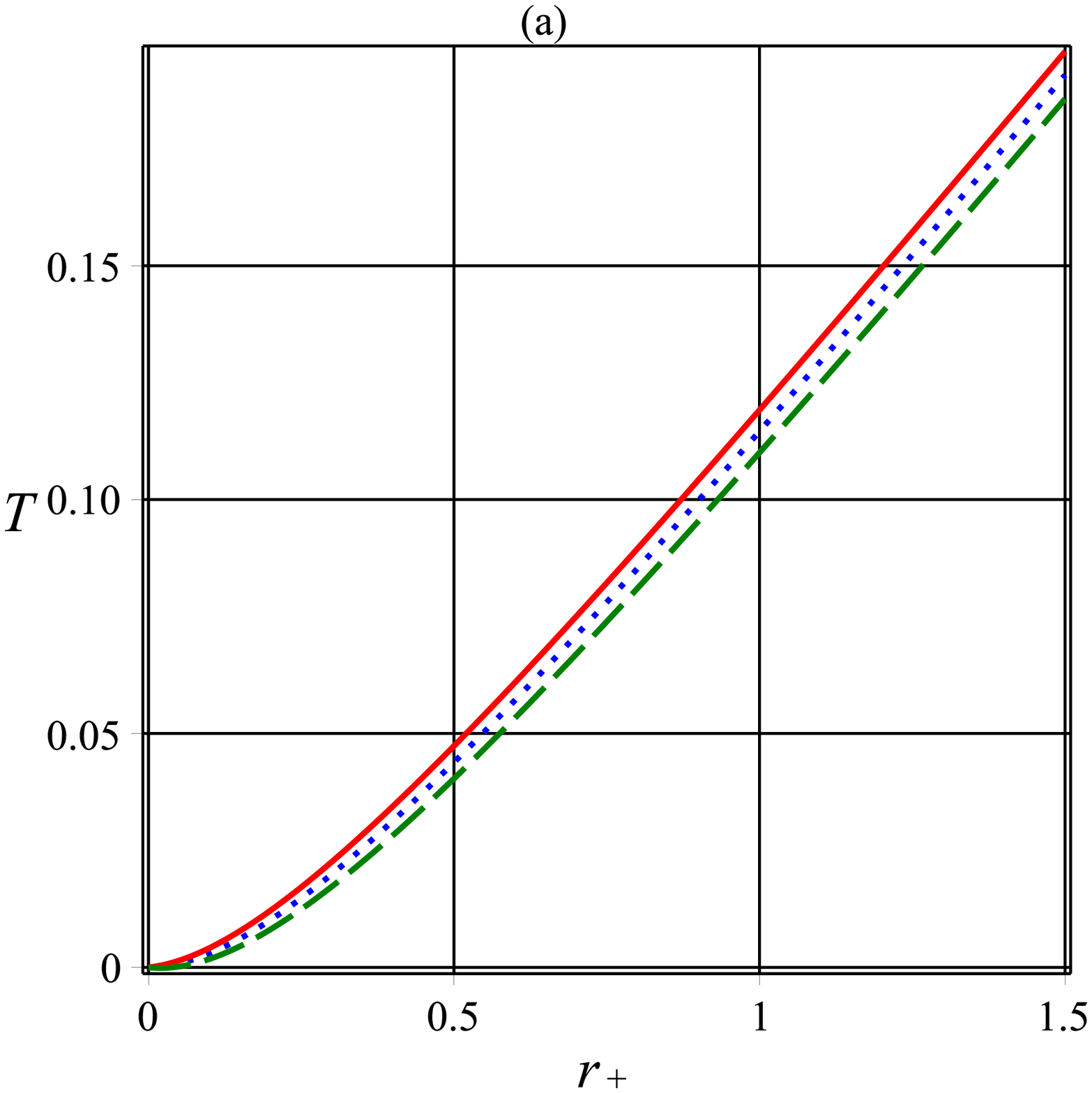}\includegraphics[width=60 mm]{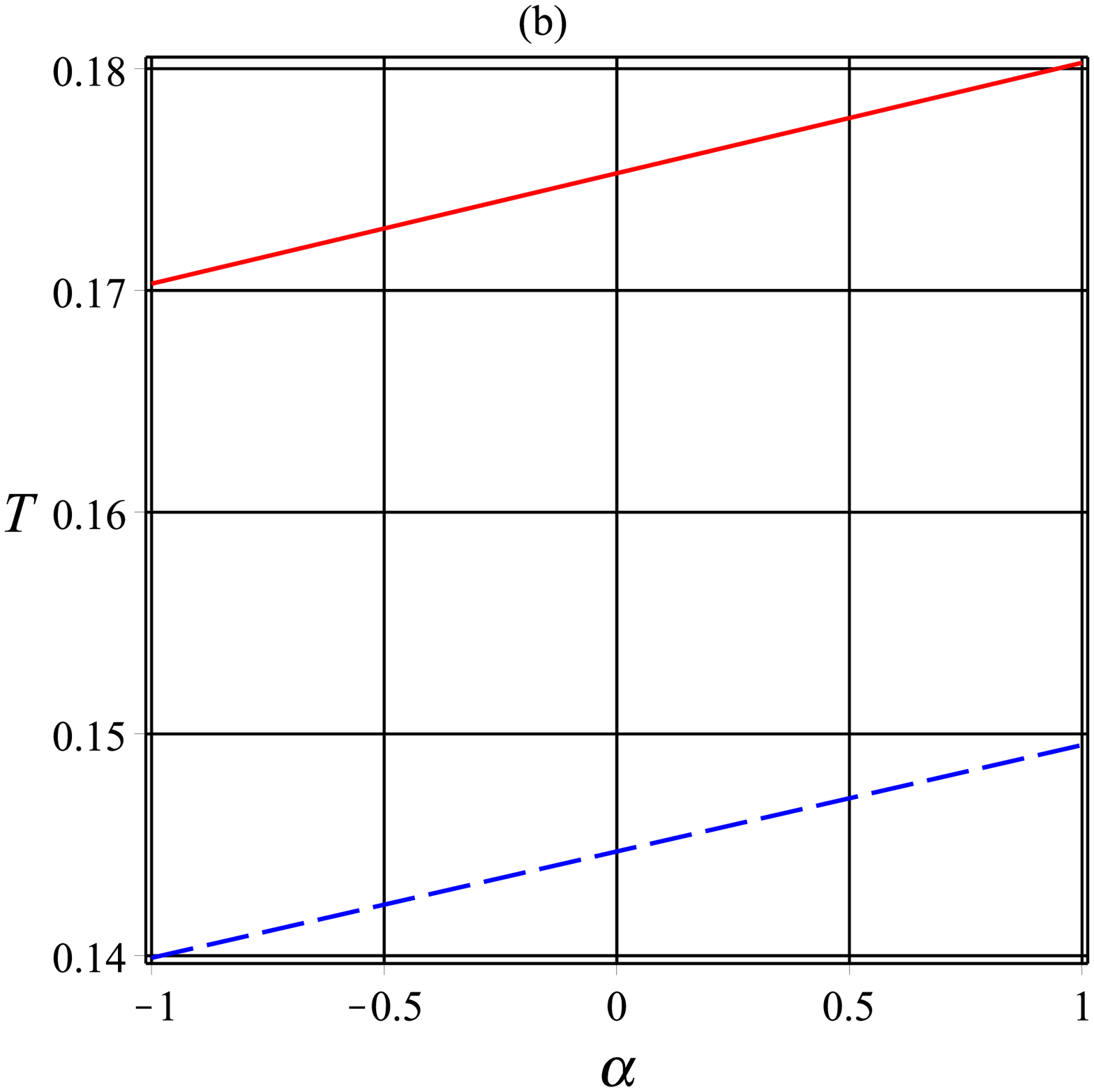}
 \end{array}$
 \end{center}
\caption{Typical behavior of the corrected Hawking temperature of uncharged hairy AdS black hole for $B=\ell=1$. (a) in terms of $r_{+}$ (b) in terms of correction coefficient.}
 \label{fig:7}
\end{figure}

In the Fig. \ref{fig:7} (a) we show shift of Hawking temperature due to the quantum correction. From the Fig. \ref{fig:7} (b) we can see that Hawking temperature is increasing function of correction coefficient. It means that Hawking temperature of uncharged hairy AdS black hole enhanced due to the thermal fluctuations.\\
From the equation (\ref{31}), the corrected physical mass for the uncharged hairy AdS black hole is,
\begin{equation}
M =\frac{3 \left( 8 \pi r_{+} - \alpha \ln (4 \pi r_{+} ) \right)^{3} }{ 64 \pi ^{2}\ell^{2} \left( 24 \pi r_{+}  - 3 \alpha \ln (4 \pi r_{+} ) + 16 \pi B \right)},
\end{equation}

\begin{figure}[h!]
 \begin{center}$
 \begin{array}{cccc}
\includegraphics[width=60 mm]{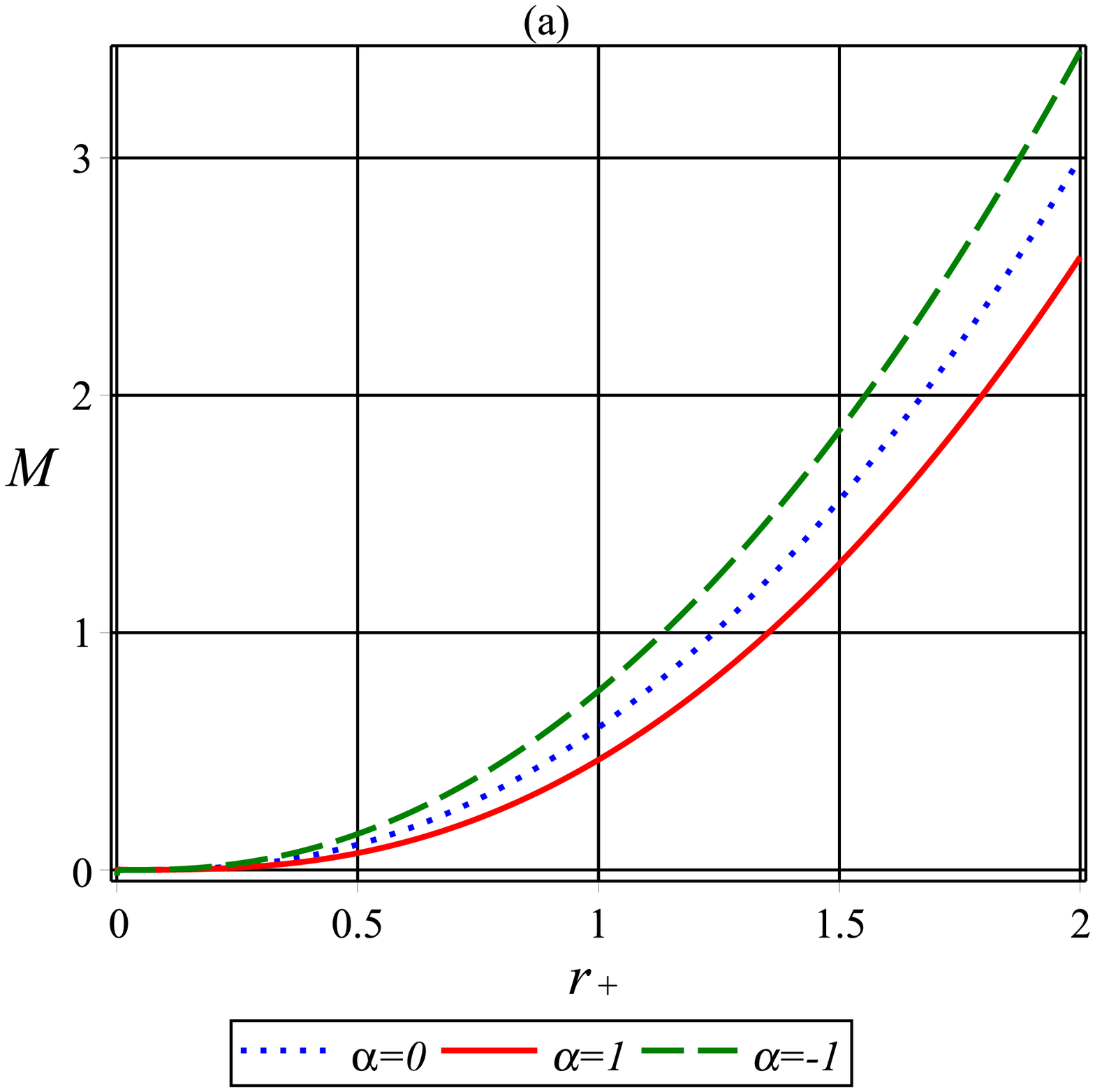}\includegraphics[width=60 mm]{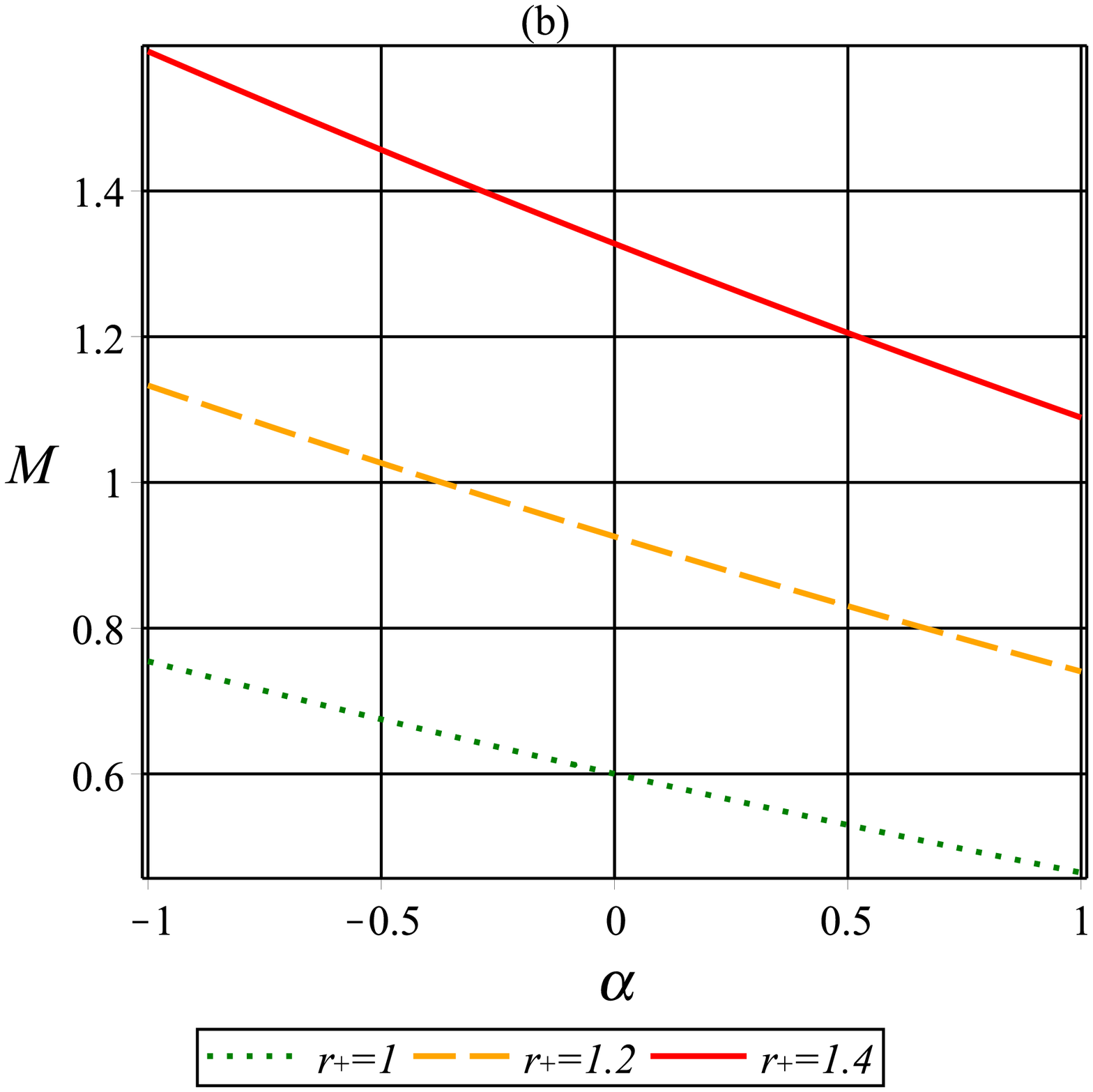}
 \end{array}$
 \end{center}
\caption{Typical behavior of uncharged hairy AdS black hole mass for $B=\ell=1$. (a) in terms of $r_{+}$ (b) in terms of correction coefficient.}
 \label{fig:8}
\end{figure}

\begin{figure}[h!]
 \begin{center}$
 \begin{array}{cccc}
\includegraphics[width=50 mm]{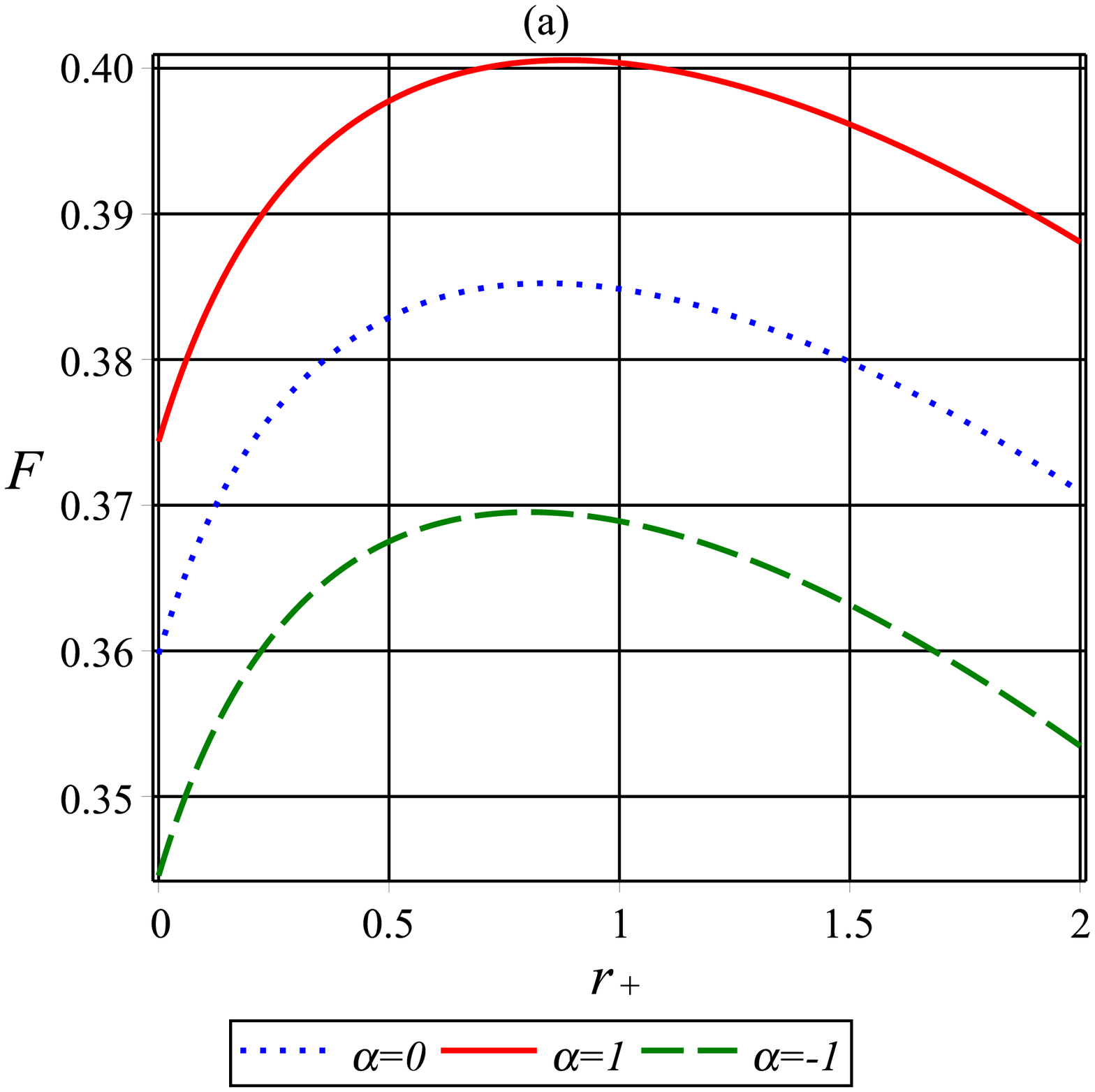}\includegraphics[width=50 mm]{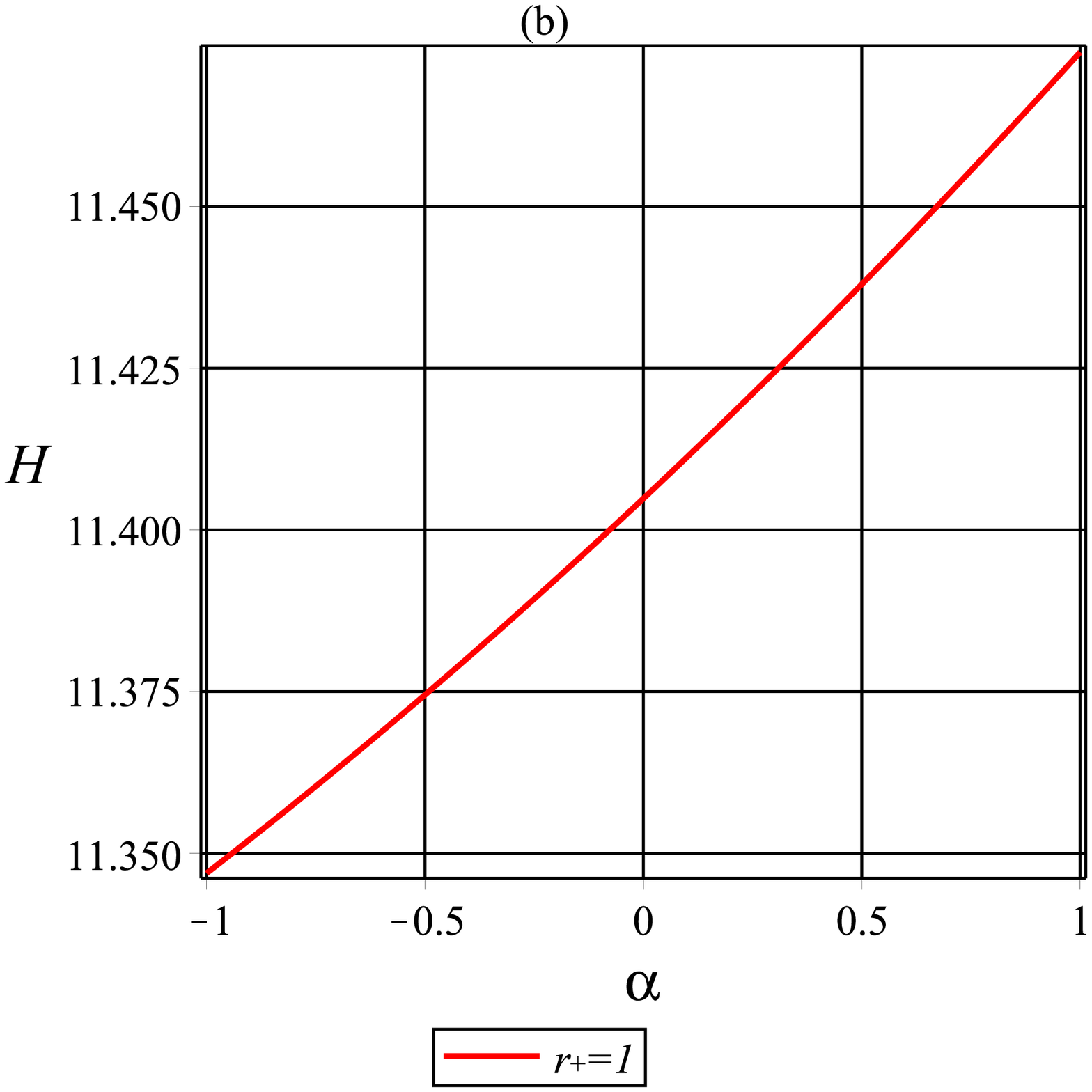}\\
\includegraphics[width=50 mm]{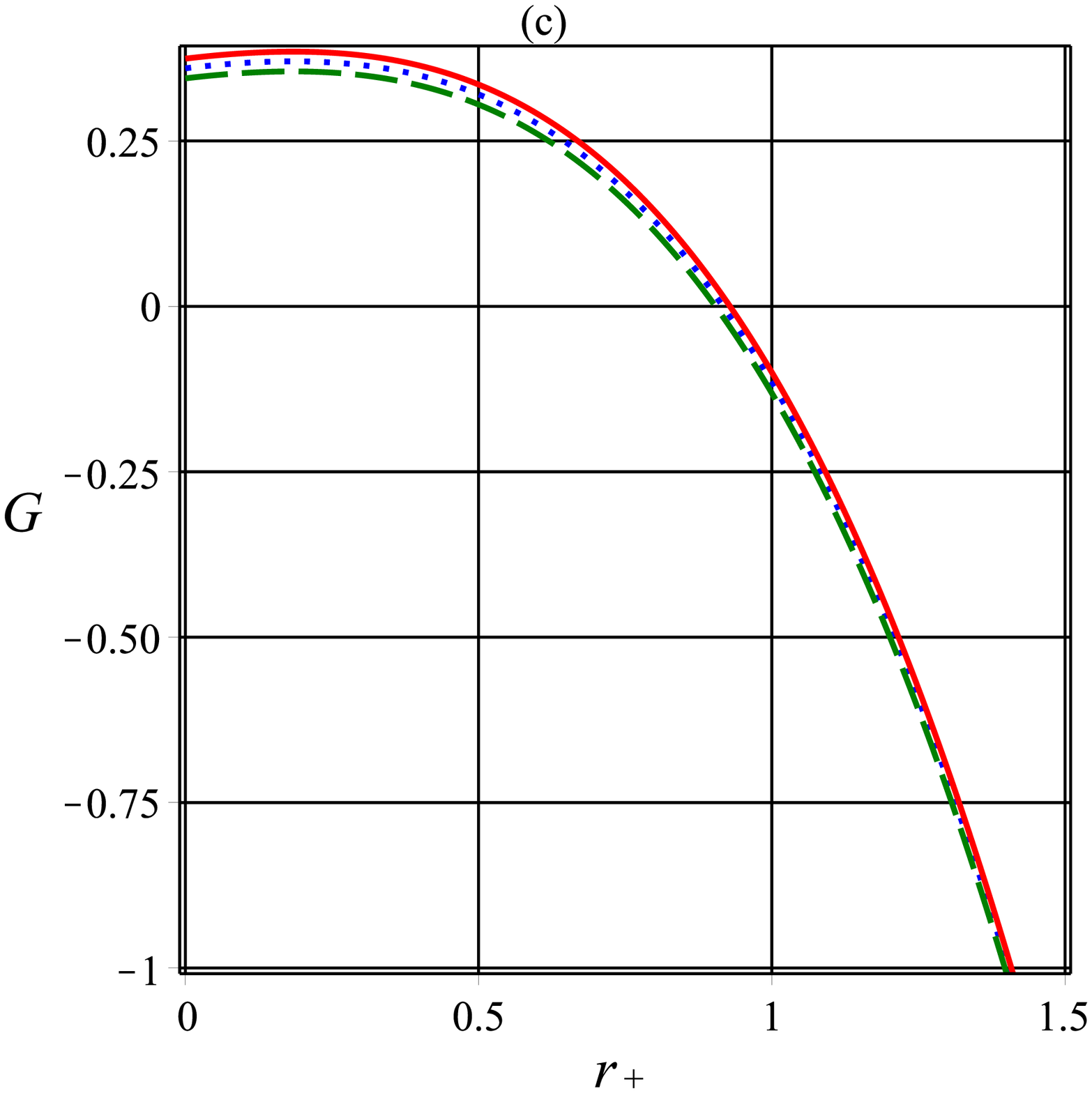}\includegraphics[width=50 mm]{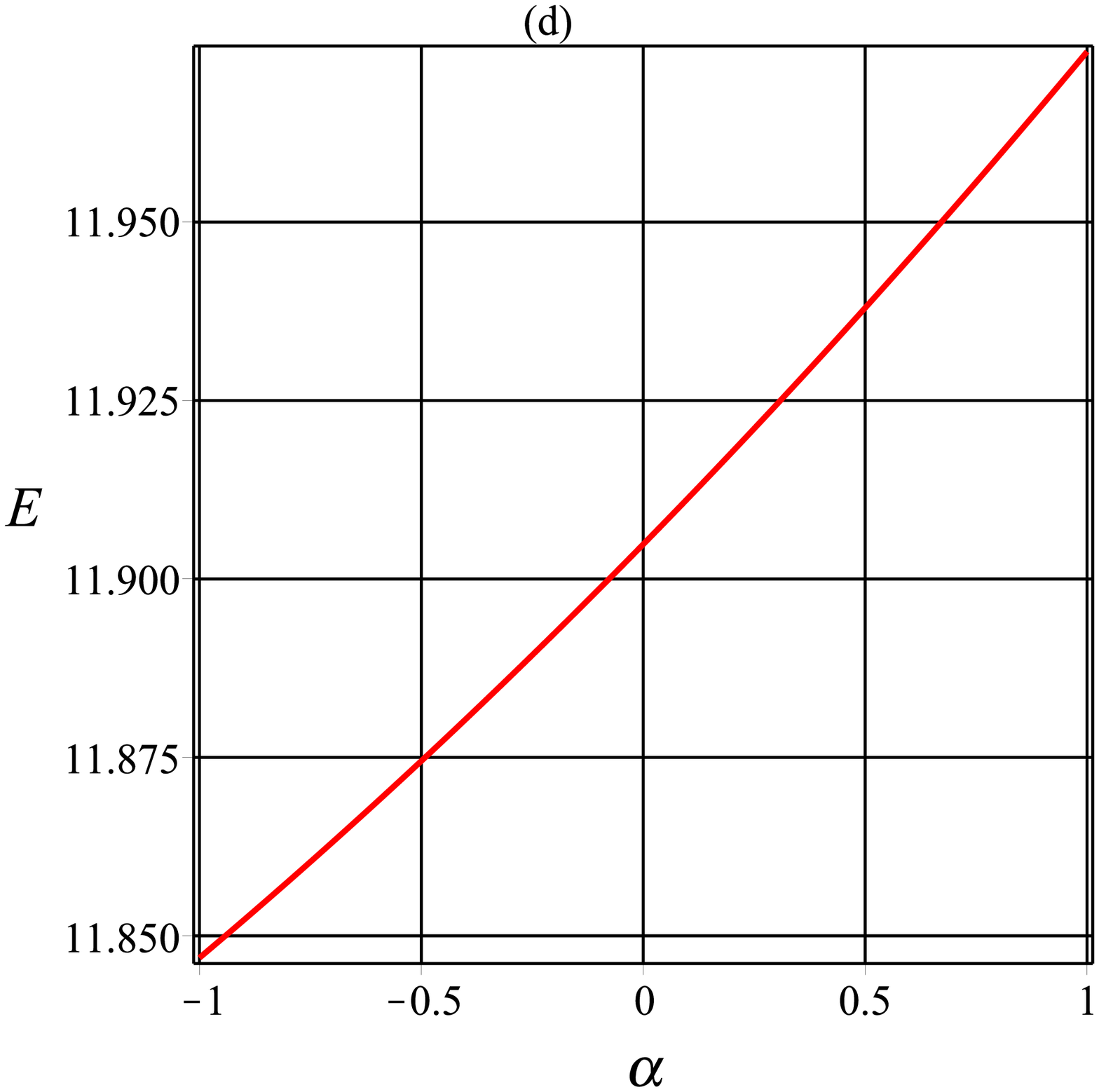}
 \end{array}$
 \end{center}
\caption{Typical behavior of (a) Helmholtz free energy (b) internal energy (c) enthalpy (d) Gibbs free energy for uncharged hairy AdS black hole with $B=\ell=1$.}
 \label{fig:9}
\end{figure}

In the Fig. \ref{fig:8} (a) we show shift of black hole mass due to the quantum correction. Fig. \ref{fig:8} (b) shows that uncharged hairy AdS black hole mass is completely decreasing function of correction parameter. It means that there is upper bound for the correction coefficient to have positive mass which is about $\alpha\approx10$ for $r_{+}\approx1$.\\
The first law of thermodynamic for such black hole given by the equation (\ref{fl}),
So, here one can obtain the Helmholtz free energy for uncharged hairy AdS black hole which is obtained by the following expression,
\begin{eqnarray}
F &=&   \frac{2 B}{(12 \pi \ell)^2} \left[  24 \pi B \ln(u) - u  + \frac{3 (8 \pi B)^2}{u} -\frac{(8 \pi B)^3}{2 u^2} \right]\nonumber\\
&+&\frac{\alpha  B}{ 3 (4 \pi \ell)^2}\left[  \frac{(\ln(u))^2}{2}  - \ln(u) + \frac{(8 \pi B)^2}{4 u^2} [1 - 2 \ln(u)] + 16 \pi B \frac{\ln(u)}{u}\right]\nonumber\\
&+&\frac{\alpha^2  B}{ 2 (4 \pi \ell)^2}\left[ \frac{(2 \pi B)}{ u^2} - \frac{1}{u}(1 + \ln(u)) + (4 \pi B)\frac{\ln(u)}{u^{2}} \right],
\end{eqnarray}
where $u=3S_{0}+8\pi B$ is defined.\\
The internal energy $ E $ can be obtained as,
\begin{eqnarray}
E &=&   \frac{2 B}{(12 \pi \ell)^2}\left[ 24 \pi B \ln(S_{0}) - u + \frac{3 (8 \pi B)^2}{u} -\frac{(8 \pi B)^3}{2 u^2} + 81 \frac{S_{0}^{3}}{ B} \left( \frac{( S_0 + 4 \pi B)}{u^{2}}\right)   \right]\nonumber\\
&+& \frac{\alpha  B}{ 6 (4 \pi \ell)^2} \left[ (\ln(u))^2 -  [ \frac{54 S_{0} }{B} (S_{0}^{2} + 4 \pi B) + ( 8 \pi B)^2] \frac{\ln(u)}{u^2} - 2 \ln(u)  + (32 \pi B )  \frac{\ln(u)}{u}\right]\nonumber\\
&+& \frac{\alpha  B}{ 6 (4 \pi \ell)^2} \left[  54 \frac{S_{0}^{2}}{ B} \left( \frac{( S_0 + 4 \pi B)}{u^{2}}\right)+ 32 \frac{( \pi B)^2}{ u^2}   \right]\nonumber\\
&+&\frac{\alpha^2  B}{ 2 (4 \pi \ell)^2}\left[ \frac{(2 \pi B)}{ u^2} - \frac{1}{u}(1 + \ln(u)) + (4 \pi B)\frac{\ln(u)}{u} - \frac{9 S_0}{B} (S_0 + 4 \pi B) \frac{\ln(S_0 )}{u^2}  \right].
\end{eqnarray}
Moreover, one can obtain the enthalpy as follow,
\begin{eqnarray}
H &=&   \frac{2 B}{(12 \pi \ell)^2}\left[ 24 \pi B \ln(S_{0}) - (3 S_0 + 9 \frac{S_{0}^{3}}{16 \pi B} + 8 \pi B)  + \frac{3 (8 \pi B)^2}{(3 S_0 + 8 \pi B)} \right]\nonumber\\
&+&\frac{2 B}{(12 \pi \ell)^2} \left[ -\frac{(8 \pi B)^3}{2 u^2} + 81 \frac{S_{0}^{3}}{ B}  \frac{( S_0 + 4 \pi B)}{u^{2}}   \right]\nonumber\\
&+&\frac{\alpha  B}{ 6 (4 \pi \ell)^2} \left[ (\ln(u))^2 -  [ \frac{54 S_{0} }{B} (S_{0}^{2} + 4 \pi B) + ( 8 \pi B)^2] \frac{\ln(u)}{u^2} - 2 \ln(u) + 32 \pi B   \frac{\ln(u)}{u}\right]\\
&+&\frac{\alpha  B}{ 6 (4 \pi \ell)^2} \left[32 \frac{( \pi B)^2}{ u^2} + 54 \frac{S_{0}^{2}}{ B}  \frac{( S_0 + 4 \pi B)}{u^{2}}   \right]\nonumber\\
&+&\frac{\alpha^2  B}{ 2 (4 \pi \ell)^2}\left[ \frac{2 \pi B}{ u^2} - \frac{1}{u}(1 + \ln(u)) + 4 \pi B \frac{\ln(u)}{u} - \frac{9 S_0}{B} (S_0 + 4 \pi B) \frac{\ln(S_0 )}{u^2}  \right].
\end{eqnarray}
Finally, the corrected Gibbs free energy is given by,
\begin{eqnarray}
G &=&   \frac{2 B}{(12 \pi \ell)^2} \left[  24 \pi B \ln(u) - (3 S_0 + 9 \frac{S_{0}^{3}}{16 \pi B} + 8 \pi B)     + \frac{3 (8 \pi B)^2}{u} -\frac{(8 \pi B)^3}{2 u^2} \right]\nonumber\\
&+& \frac{\alpha  B}{ 3 (4 \pi \ell)^2}\left[  \frac{(\ln(u))^2}{2}  - \ln(u) + \frac{(8 \pi B)^2}{4 u^2} [1 - 2 \ln(u)] + 16 \pi B \frac{\ln(u)}{u}\right]\nonumber\\
&+& \frac{\alpha^2  B}{ 2 (4 \pi \ell)^2}\left[ \frac{(2 \pi B)}{ u^2} - \frac{1}{u}(1 + \ln(u)) + (4 \pi B)\frac{\ln(u)}{u} \right].
\end{eqnarray}
Graphically, we analyze above quantities and obtain quantum correction effects (see Fig. \ref{fig:9}).

In the Fig. \ref{fig:9} (a) we draw Helmholtz free energy in terms of event horizon radius and see that there is a maximum which is corresponding to the black hole stability. This maximum affected by quantum corrections which is illustrated by the Fig. \ref{fig:9} (a). It means that Helmholtz free energy is increased by quantum corrections.\\
Figs. \ref{fig:9} (b) and (d) show that effect of correction on the internal energy and enthalpy is infinitesimal however both of them are increasing function of $\alpha$. Fig. \ref{fig:9} (c) show that effect of quantum correction on the Gibbs energy is important when $r_{+}$ is small. It means that Gibbs free energy of large black hole sense no any effect of thermal fluctuations.\\
We will discuss the critical points and the stability of the   uncharged hairy AdS black hole in the next section.

\subsection{Stability of uncharged hairy AdS black hole}
The critical point $(r_{+,c}, P_{c}, T_{c})$ for uncharged hairy AdS black hole corresponding to the phase transition can be obtained by the condition (\ref{Con}), which yields to the following critical point,
\begin{equation}
r_{+,c} = \frac{-  1  \pm \sqrt{65} }{8} B.
\end{equation}
Now, we want to discuss the global stability condition for the corresponding system. The graphical analysis of the Gibbs free energy  in the case of quantum corrected entropy and  the corrected temperature for the uncharged hairy AdS black hole can  see in Fig. \ref{fig:9} (c) where we fix constant $ B $ and $ \ell $. In Fig. \ref{fig:9} (c),  we observe that correction terms are a few  effects on the Gibbs free energy. But  the Gibbs free energy will have global stability for enough large event horizon radius. For the selected value of model parameter $B=\ell=1$, we have global stability ($G<0$) if $r_{+}>0.9$ which means $M>1$ (see Fig. \ref{fig:1} (b)).\\
In order to discuss the local stability of the black hole we need to calculate the specific heat at constant pressure which is given by,
\begin{equation}
C_{P} =  \frac{ 32 \pi ^{2} r_{+}^{2}  (3 r_{+} + 2 B) [8 \pi r_{+}^{2} + 8 \pi r_{+} B + \alpha (r_{+} + B)]}{\left( 24 \pi r_{+}^{3} +24 \pi B r_{+}^{2} + 2 B^{2} (8 \pi r_{+} +\alpha ) + \alpha B r_{+}\right) (8 \pi r_{+}  - \alpha) }.
\end{equation}

\begin{figure}[h!]
 \begin{center}$
 \begin{array}{cccc}
\includegraphics[width=60 mm]{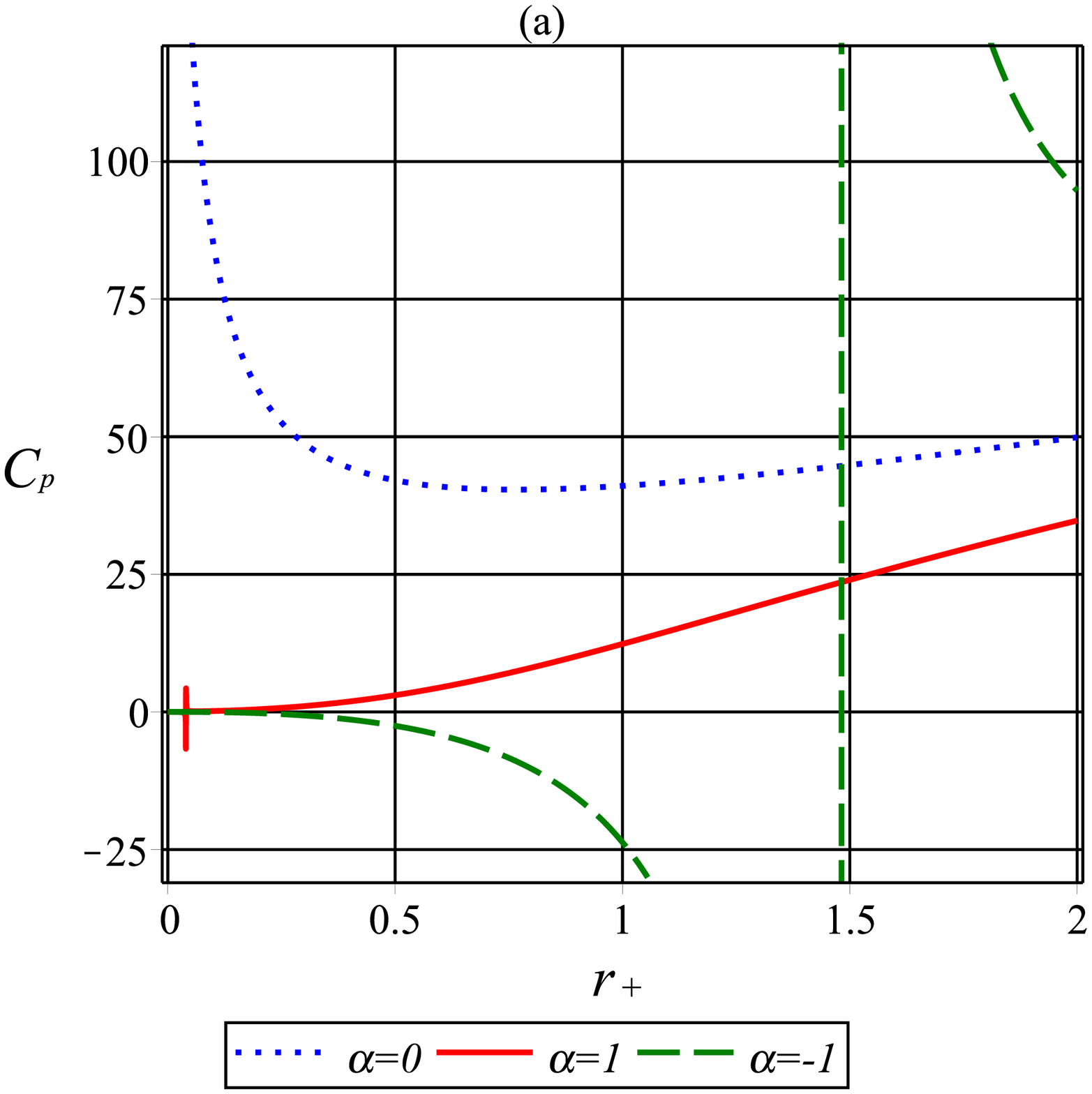}\includegraphics[width=60 mm]{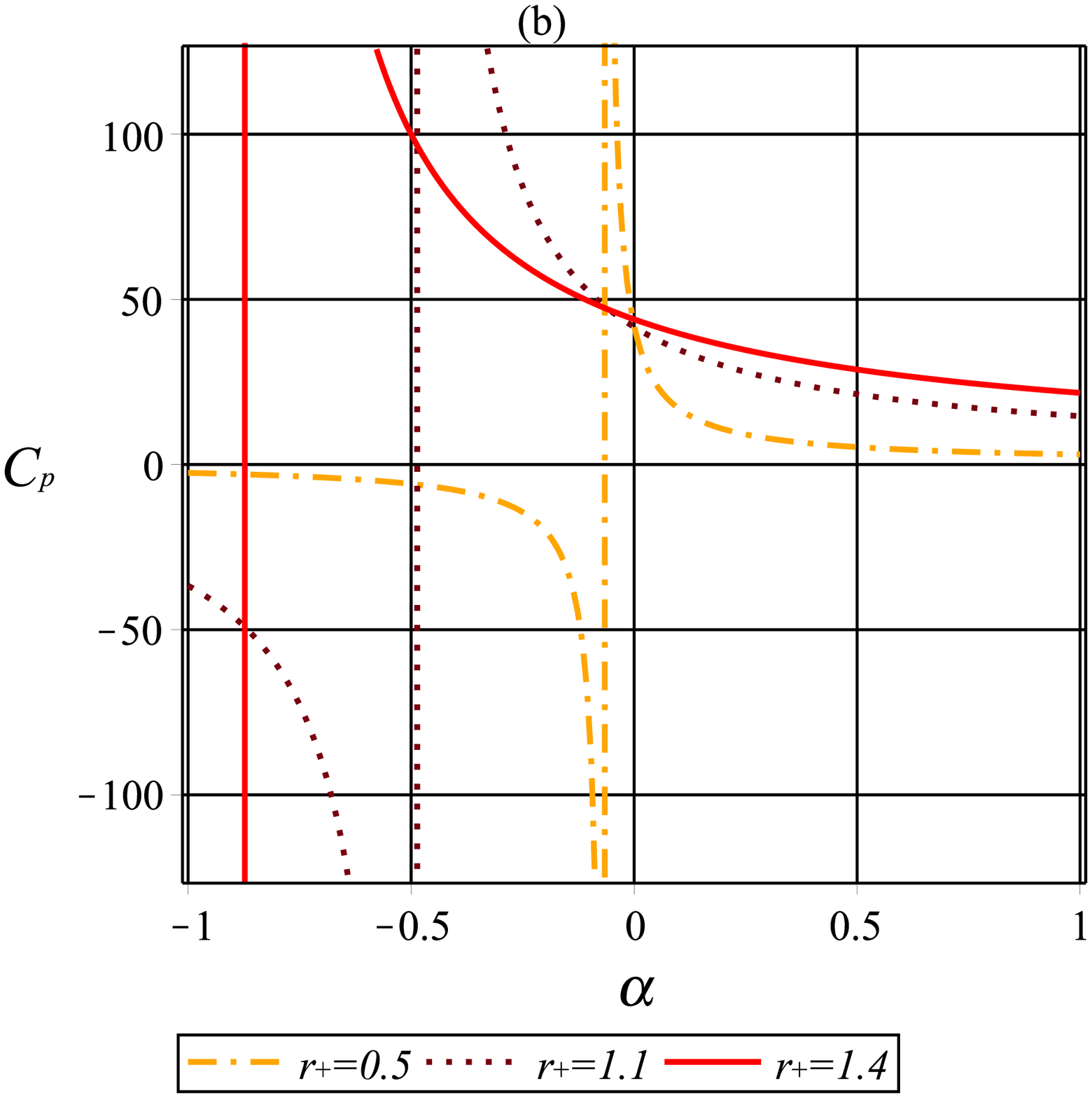}
 \end{array}$
 \end{center}
\caption{Typical behavior of specific heat at constant pressure of uncharged hairy AdS black hole with $B=\ell=1$.}
 \label{fig:10}
\end{figure}

We draw $C_P$ in terms of $ r_{+}$  with  constant values of $B$ parameter and quantum correction coefficient  $\alpha$ for uncharged hairy AdS black hole in Fig. \ref{fig:10}. Both $r_{+}$ and $\alpha$ dependence figures (Fig. \ref{fig:10} (a) and (b)) show that ignoring thermal fluctuation ($\alpha=0$) yields to completely stable black hole. Quantum correction with positive $\alpha$ reduces value of specific heat but still it is positive and black hole is stable. However, negative corrected coefficient yields to stable/unstable phase transition. Dashed green line of the Fig. \ref{fig:10} (a) show phase transition at about $r_{+}\approx1.5$. We can see that if $r_{+}<1.5$ (by choosing $B=\ell=1$), which means $M<1.5$ (see Fig. \ref{fig:1} (b)) then black hole is in unstable phase, while if $r_{+}>1.5$ ($M>1.5$) the black hole is thermodynamically stable. The important consequence of quantum correction is that the black hole is unstable for some values of negative $ \alpha $ and stable for positive $\alpha$. We can see that there are some stable region $ r_{+} \geq r_{m}  $ corresponding to $ \alpha \geq  0 $. Furthermore, $ r_{+} = r_{m} $ is a minimum value of the black hole horizon radius where the phase transition happen.\\
The heat capacity at constant volume is,
\begin{equation}
C_{V} = \left[ \frac{32 \pi^{2} r_{+}^{2} (r_{+} + B) (3 r_{+} + 2 B)}{24 \pi r_{+}^{3} +24 \pi r_{+}^{2} B + \alpha B (r_{+} +2 B)}  \right] ,
\end{equation}

\begin{figure}[h!]
 \begin{center}$
 \begin{array}{cccc}
\includegraphics[width=65 mm]{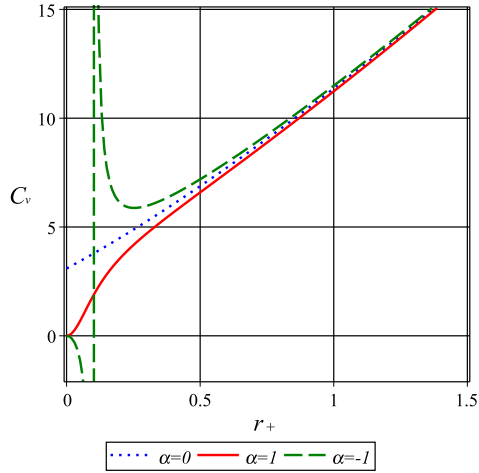}
 \end{array}$
 \end{center}
\caption{Typical behavior of specific heat at constant volume of uncharged hairy AdS black hole with $B=\ell=1$.}
 \label{fig:11}
\end{figure}

We draw $C_V $ in terms of $r_{+}$  with  different values of quantum correction coefficient  $\alpha$ for uncharged hairy AdS black hole in Fig. \ref{fig:11}.
We can see effect of quantum corrections are important when event horizon radius is small. For the large black hole, thermal fluctuations have no any important effect on the specific heat at constant volume.\\
The local stability equilibrium condition for a thermodynamic system reads as,
\begin{equation}
C_P \geq C_V \geq 0,
\end{equation}
which yields,
\begin{equation}
(8 \pi r_{+} + \alpha) [8 \pi r_{+}^{2} + 8 \pi r_{+} B + \alpha (r_{+} + B)] \geq 3 (r_{+}  + B) (8 \pi r_{+} + \alpha).
\end{equation}
Therefore, we can find,
\begin{equation}
\alpha \geq 0,
\end{equation}
as expected. It is completely agree with choosing $\alpha=1$ \cite{59}.

\section{Corrected thermodynamics for conformally  dressed AdS black hole }
If  we assume $ \beta = -\frac{B^{2}}{ \ell^{2} }$,  in the equation (\ref{5}), we will have the conformally  dressed AdS black hole solution.
In this section, we will use the corrections to the entropy and the temperature of a conformally  dressed AdS black hole to obtain an  explicit expression for various thermodynamic quantities.
Then, we will use these explicit values to study phase transition of this system. So,
\begin{equation}\label{50}
f(r) = \frac{r^{2}}{\ell^{2}} - 3 \frac{B^{2}}{\ell^{2}} - 2 \frac{B^{3}}{\ell^{2} r}.
\end{equation}
Because of $ \beta = - \frac{M}{3} $ the mass will be,
\begin{equation}\label{51}
M =\frac{3 r^{2}_{+} }{4 \ell^{2}} = 2 \pi P r^{2}_{+} ,
\end{equation}
Now, exploiting relations (\ref{50}) and (\ref{51}), the Hawking temperature of the event horizon can be calculated by
\begin{equation}
T_{0} =  \frac{3 S}{32 \pi^{2} \ell^{2} },
\end{equation}
The  entropy density of conformally  dressed AdS black hole is given by (\ref{12}). Hence, the first-order corrected entropy and temperature  for the conformally  dressed AdS black hole is computed as
\begin{equation}
S= 4 \pi r_{+} - \frac{\alpha}{2} \ln{ \left( 4 \pi r_{+}\right) }
\end{equation}
and,
\begin{equation}
T= \frac{3}{(8 \pi  \ell)^{2}} (8 \pi r_{+} + \alpha ),
\end{equation}
It is clear that the effects of quantum correction (with positive $\alpha$) is increasing the value of temperature and vice versa.\\
From equation (\ref{51}), the corrected physical mass for the conformally  dressed AdS  black hole is,
\begin{equation}
M = \frac{3}{4 (8 \pi \ell)^{2}} \left( 64 \pi^{2}  r^{2}_{+}  +\alpha^{2} \left(\ln( 4 \pi r_{+}) \right)^{2} - 16 \alpha \pi r_{+} \ln(4 \pi r_{+} )  \right),
\end{equation}
We find that effect of thermal fluctuations are infinitesimal on the black hole mass.\\
The thermodynamic first law of the black hole reads,
\begin{equation}
dM = T dS + V d P
\end{equation}
We can obtain the Helmholtz free energy for the conformally dressed AdS black hole as follow,
\begin{equation}
F = - \frac{3 r^{2}_{+} }{4 \ell^{2}} + \left( \frac{3 \alpha r_{+} }{(8 \pi \ell)^{2}} \right) \left( 4 \pi \ln(4 \pi r_{+}) -1\right).
\end{equation}

\begin{figure}[h!]
 \begin{center}$
 \begin{array}{cccc}
\includegraphics[width=50 mm]{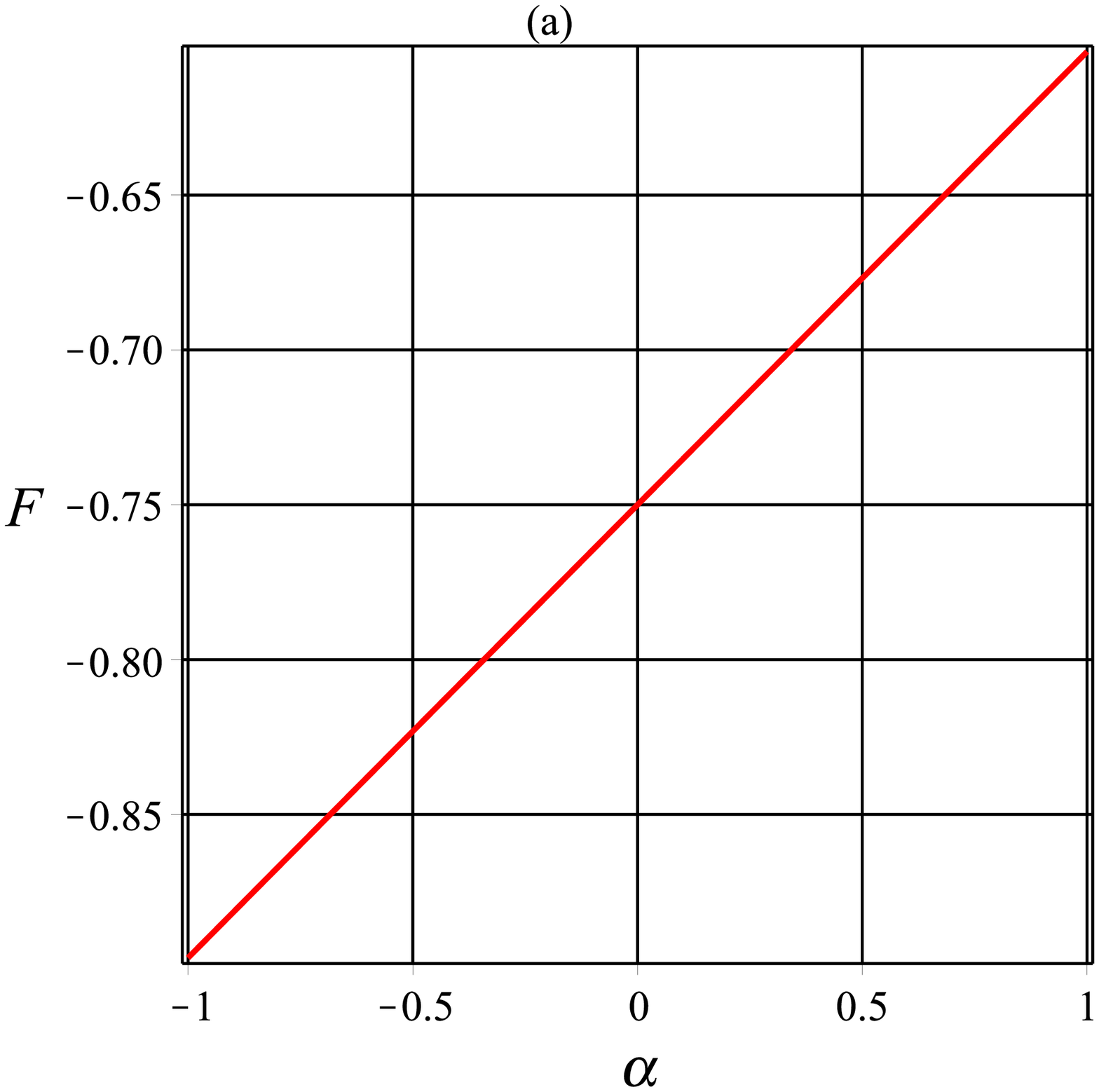}\includegraphics[width=50 mm]{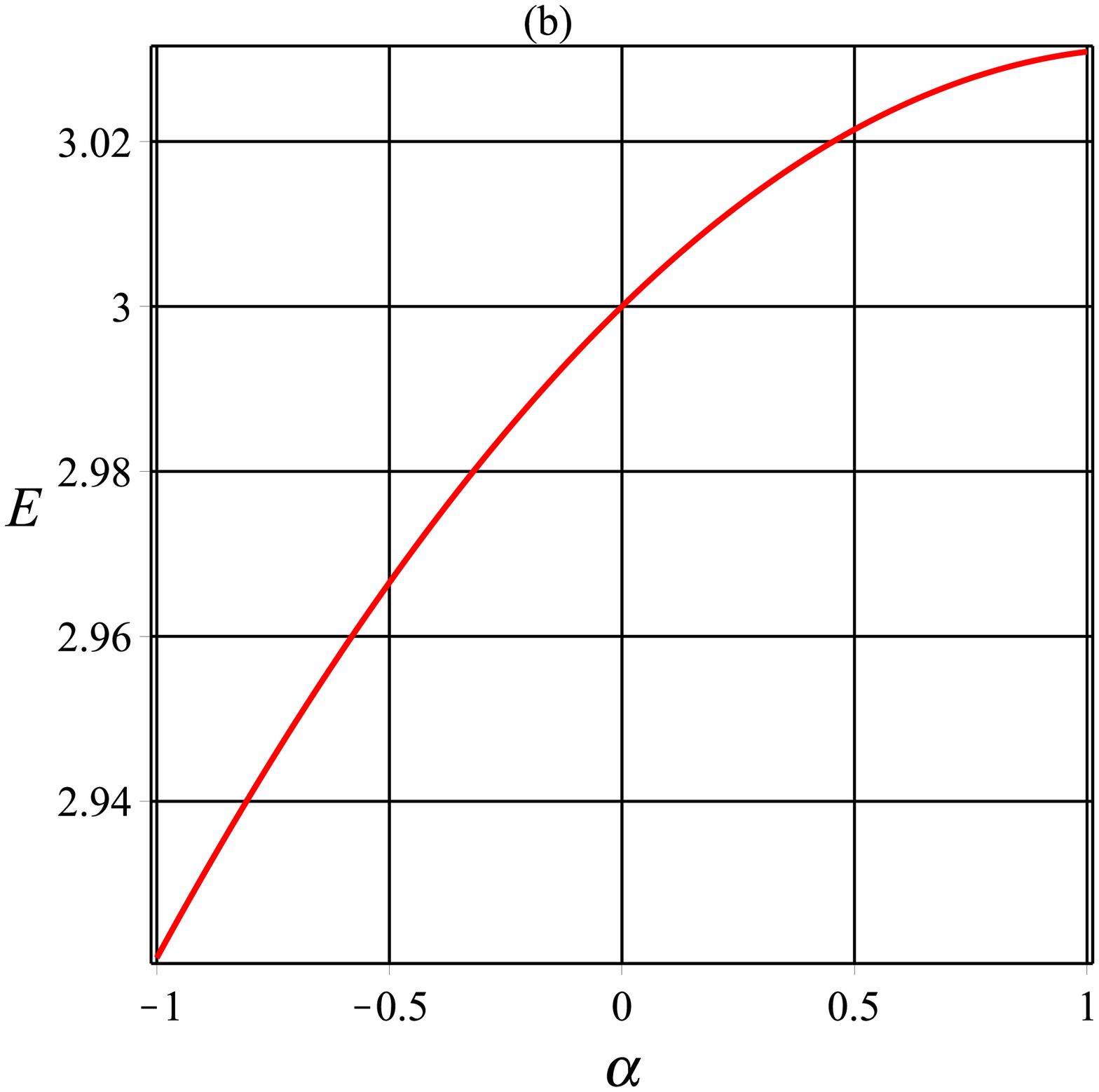}\\
\includegraphics[width=50 mm]{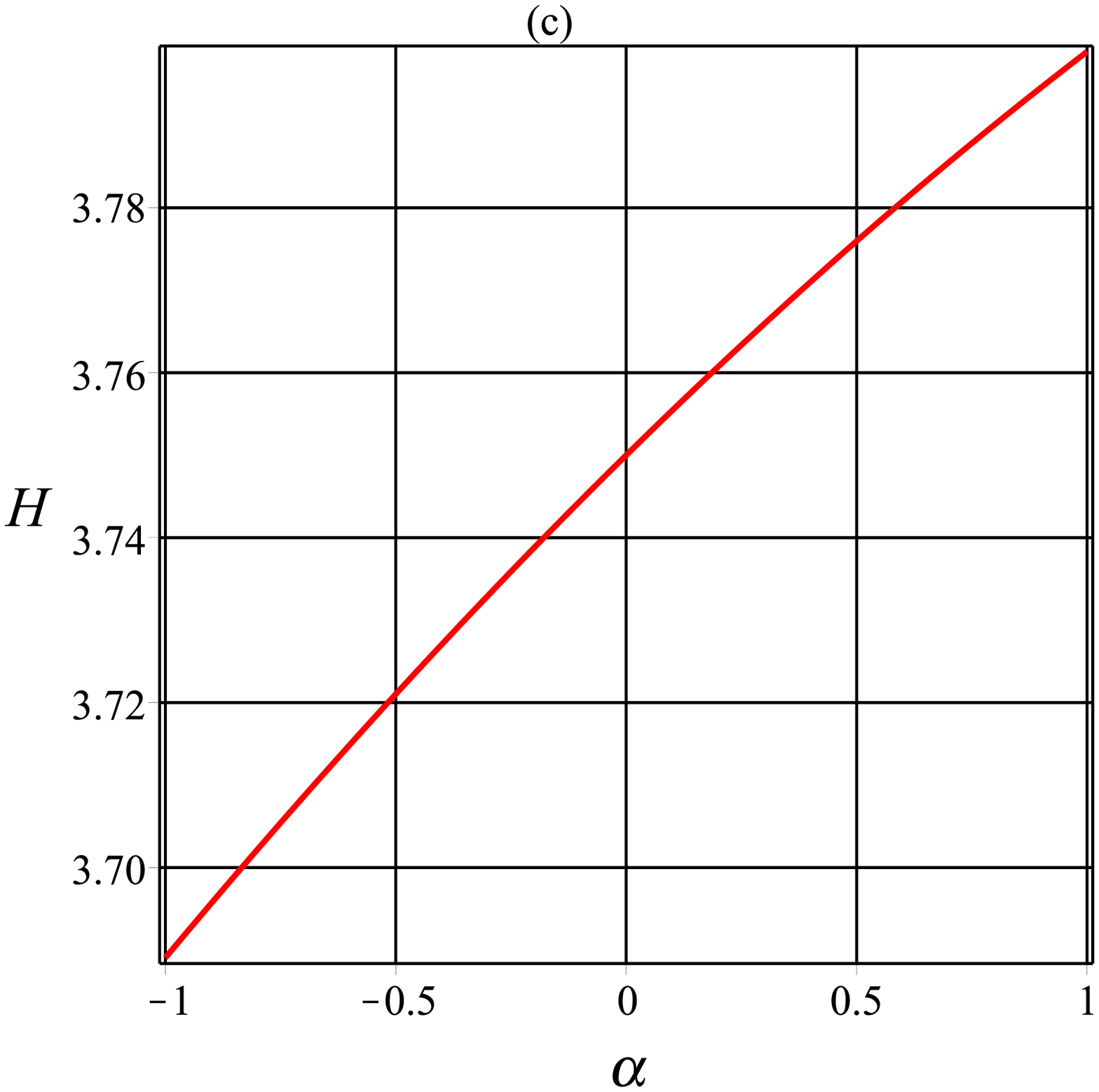}\includegraphics[width=50 mm]{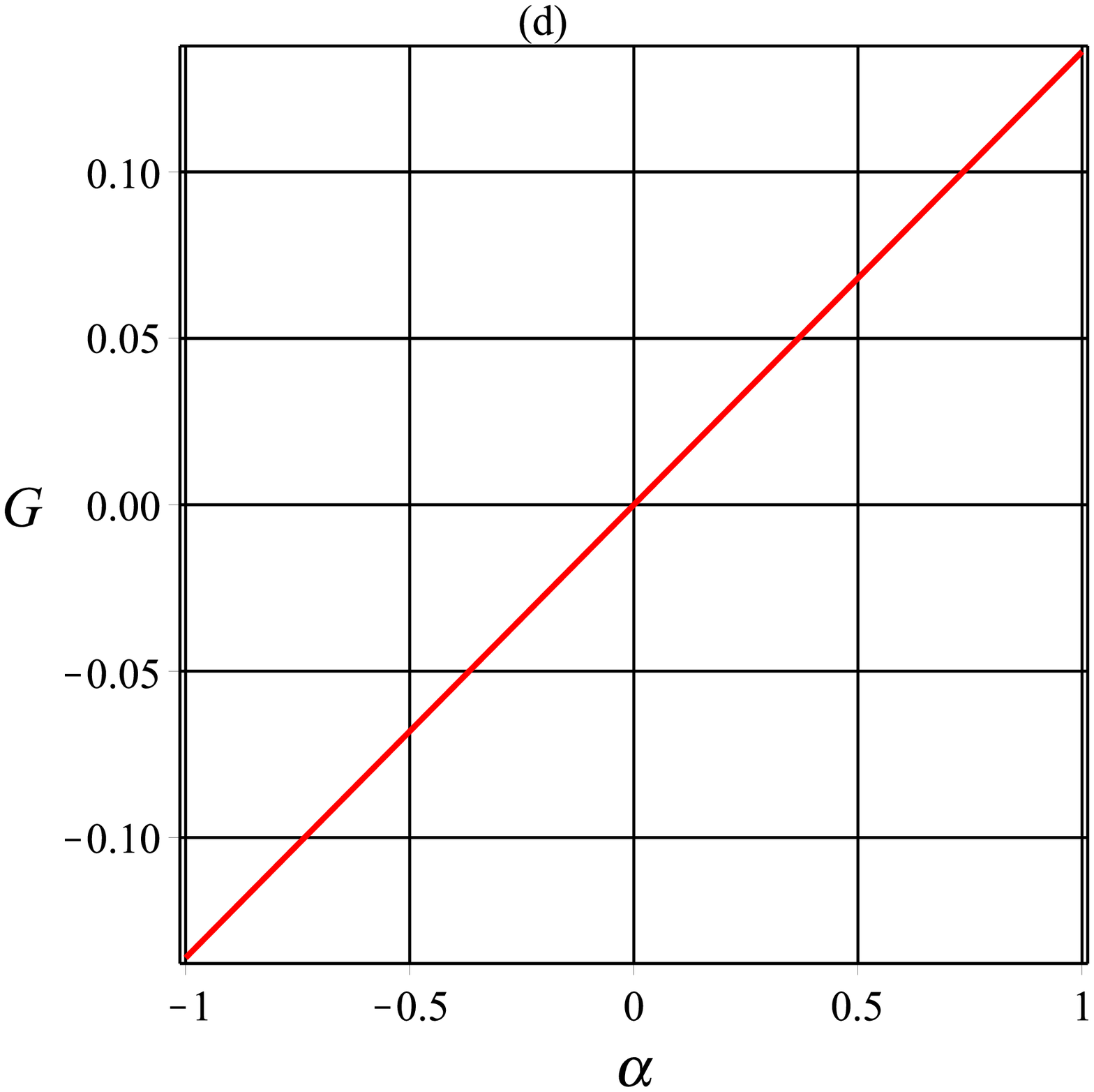}
 \end{array}$
 \end{center}
\caption{Typical behavior of (a) Helmholtz free energy (b) internal energy (c) enthalpy (d) Gibbs free energy for conformally dressed AdS black hole with $r_{+}=\ell=1$.}
 \label{fig:12}
\end{figure}

The effect of first - order correction to the Helmholtz free energy can be seen in Fig. \ref{fig:12} (a). It can be seen that with increases $ \alpha $ parameter   the Helmholtz free energy is increased.\\
The internal energy is calculated as,
\begin{equation}
E =  \frac{3 r^{2}_{+} }{ \ell^{2}} + \left( \frac{3 \alpha r_{+} }{(8 \pi \ell)^{2}} \right) \left( 4 \pi -1 \right) - \alpha^{2}  \left( \frac{3 }{ 2 (8 \pi \ell)^{2}} \right)
\ln(4 \pi r_{+}),
\end{equation}
We have plotted the internal energy in Fig. \ref{fig:12} (b). We see that the internal energy increased by any value of $\alpha$. We can see the same behavior for the enthalpy in Fig. \ref{fig:12} (c) given by
\begin{equation}
H =  \frac{15 r^{2}_{+} }{ 4 \ell^{2}} + \left( \frac{3 \alpha r_{+} }{(8 \pi \ell)^{2}} \right) \left( 4 \pi -1 \right) - \alpha^{2}  \left( \frac{3 }{ 2 (8 \pi \ell)^{2}} \right)
\ln(4 \pi r_{+}).
\end{equation}
We can obtain the Gibbs free energy as the  following expression,
\begin{equation}
G =   \frac{3 \alpha r_{+} }{(8 \pi \ell)^{2}}  \left[ 4 \pi \ln(4 \pi r_{+}) -1\right],
\end{equation}
which its behavior with correction parameter illustrated by Fig. \ref{fig:12} (d). We will discuss the critical points and the stability of the conformally  dressed AdS  black hole in the next subsection.

\subsection{Phase transition for conformally  dressed AdS black bole}

We see in Fig. \ref{fig:12} (d), the graphical analysis of the Gibbs free energy in the case of quantum corrected entropy and the temperature for conformally  dressed AdS black hole.  We observe that the correction terms increase the Gibbs free energy for such black hole when $ \alpha $ is positive.  So, the Gibbs free energy in this case,  has not global stability. Also, we can  see that the Gibbs free energy will have global stability for negative $\alpha$.
The specific heat at constant pressure with  corrected entropy and corrected temperature can  obtain as follows,
\begin{equation}
C = \frac{(8 \pi r_{+} - \alpha  ) (8 \pi r_{+} + \alpha  )}{16 \pi r_{+}}.
\end{equation}

\begin{figure}[h!]
 \begin{center}$
 \begin{array}{cccc}
\includegraphics[width=60 mm]{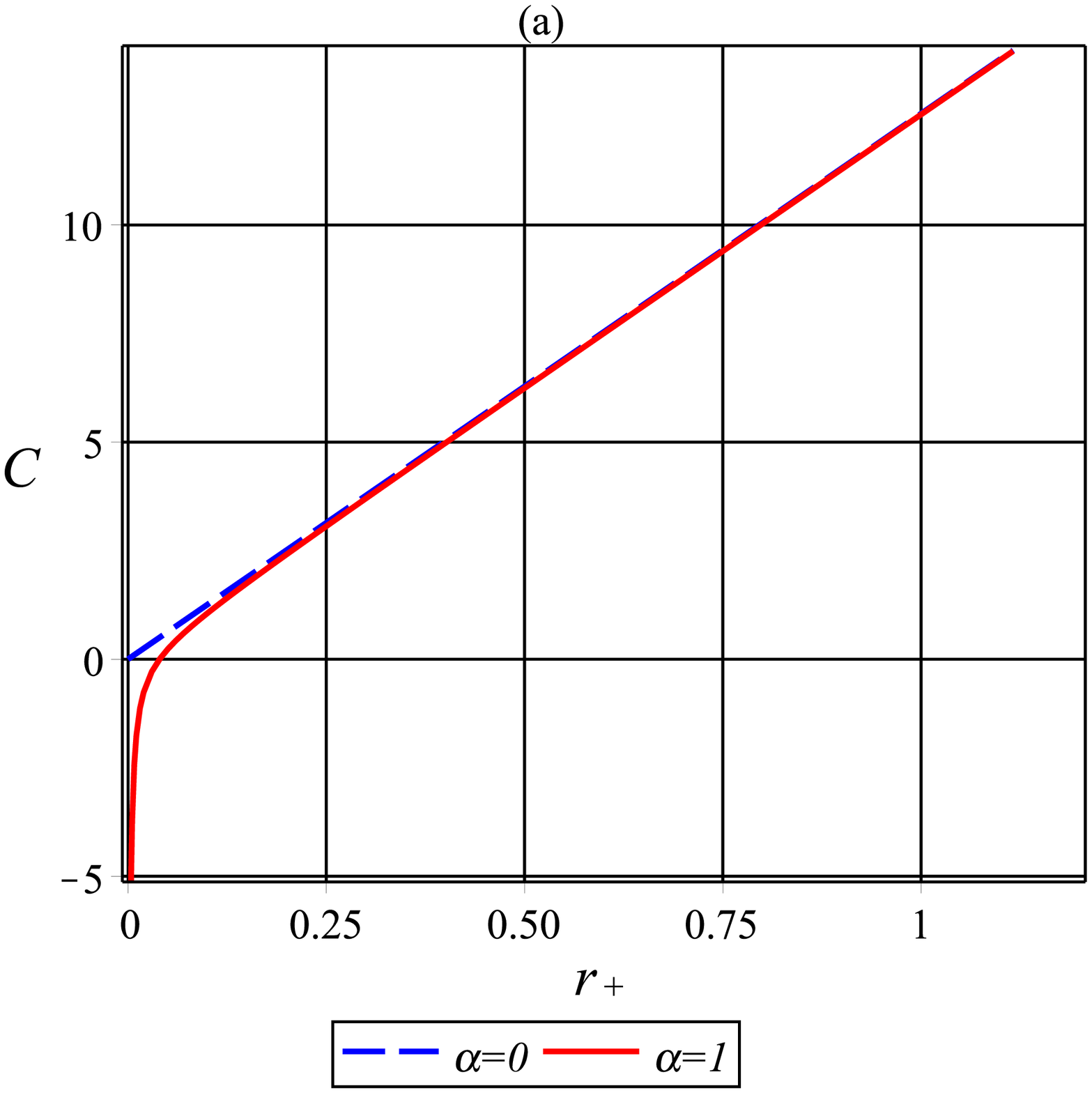}\includegraphics[width=60 mm]{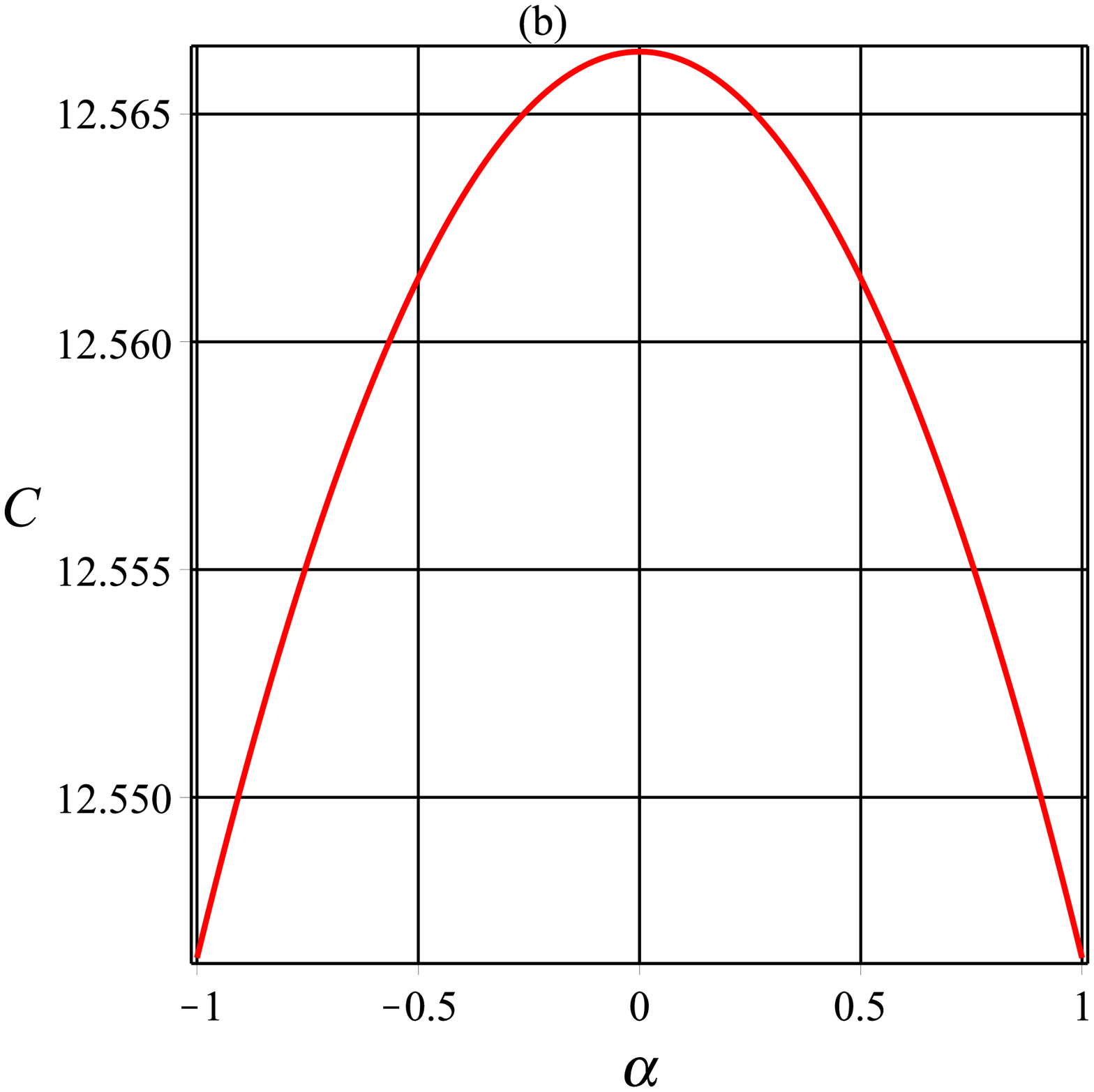}
 \end{array}$
 \end{center}
\caption{Typical behavior of specific heat at constant pressure of conformally  dressed AdS black hole with $\ell=1$.}
 \label{fig:13}
\end{figure}

In Fig.  \ref{fig:13} (a), we observe the behavior of the heat capacity at constant pressure for some values of the  $ \alpha $ parameter. It is clear that negative and positive $\alpha$ yields to similar result.
In this diagram, we can see that the small values of $ \alpha $   have a few effect  on the graph ($ \alpha \approx0 $).  But  by choosing $ \alpha > 0 $, the corrected heat capacity is decreasing.
For the large of $ \alpha $  with positive values, the effect of the corrected   heat capacity with the negative $ \alpha $ is the same.\\
Fig.  \ref{fig:13} (b) show that the black hole may be instable for small $r_{+}$, which means that thermal fluctuations is important when the size of black hole decreased due to the Hawking radiation. 

 \section{Conclusion}
In this paper, we have analyzed the thermodynamics of hairy black hole that contains a scalar field coupled minimally or nonminimally to gravity.
First of all,  we imply different cases of the charged hairy black holes.
We have studied effects of the entropy and temperature with logarithmic corrected terms on the thermodynamics of hairy black holes. Considering the corrections to the entropy and temperature, specific heat and thermodynamical quantities  were calculated for charged BTZ, uncharged hairy AdS,  and  conformally dressed AdS black holes.
Besides, we calculate internal energy, enthalpy, Helmholtz energy and Gibbs free energy for the mentioned black holes and analyze the effects of corrected entropy and temperature.\\
Critical values of event horizon radii for phase transitions are shown to was shifted due to the corrections of entropy and temperature.
This shifting is also indicated by physical mass,  specific heat and enthalpy for charged BTZ  black hole.\\
Our main work is finding the effect of corrections on the thermodynamics quantities.
Corrections exist for any black hole, but they are important for a small black holes and negligible for the large black holes. The advantage of a charged BTZ black hole  is its holographic picture, which is a van der Waals fluid. We have shown that in the presence of corrections there is still a van der Waals fluid as a dual picture. Only we should fix the black hole charge and $\ell$
which corresponds to the electric charges of a van der Waals fluid. We obtained some thermodynamics quantities like Gibbs and Helmholtz free energies and showed that corrections don't  have important effects on the large black hole.\\
On the other hand, for the small black holes, they are important and have a crucial role. Using specific heat, we found that corrections reduced stable regions of the black hole.
However,  there are enough stable regions to see quantum gravity effects before the phase transition of a van der Waals fluid.  This means that there is a minimum radius, which we call the critical radius, where a black hole is stable in the presence of corrections, and in this region, a  charged BTZ black hole in the presence of
corrections is a dual of the van der Waals fluid.\\
It is still possible to find effect of such logarithmic correction on a Hyperscaling violation black hole background \cite{60} and discuss stability conditions.

%%%%%%%%%%%%%%%%%%%%%%%%%%%%%%%%%%%%%%%%%

\begin{thebibliography}{}
\bibitem{1} S. Deser, R. Jackiw, and S. Templeton, Annals of Physics,  \textbf{140} 372 (1982).
\bibitem{2} S. Deser, R. Jackiw, and S. Templeton, Phys. Rev. Lett. \textbf{48} 975 (1982).
\bibitem{3} S. Deser and R. Jackiw, Annals of Physics,\textbf{153} 405 (1984).
\bibitem{4} S. Giddings, J. Abbott, and K. Kuchar, Gen. Relativ. Gravitation \textbf{16} 751 (1984).
\bibitem{5} M. M. Caldarelli, G. Gognola, D. Klemm, Class. Quant. Grav. \textbf{17} 399 (2000).
\bibitem{6} C. Martinez, C. Teitelboim, and J. Zanelli, Phys. Rev. D, \textbf{61} 104013 (2000).
\bibitem{7} W. Xu and L. Zhao, Phys. Rev. D, \textbf{87} 124008 (2013).
\bibitem{8} L. Zhao, W. Xu, and B. Zhu, Commun. Theor. Phys. \textbf{61} 475(2014).
\bibitem{9} J. Sadeghi, B. Pourhassan, H. Farahani, Commun. Theor. Phys.  \textbf{62} 358 (2014).
\bibitem{9-1} B. Pourhassan, Modern Phys Lett A \textbf{31} 1650057 (2016).
\bibitem{10} A. Belhaj, M. Chabab, H. EL Moumni, M.B. Sedra,   Int. J. Geom. Meth. Mod. Phys. \textbf{12} 1550017 (2014).
\bibitem{11} J. Sadeghi and H. Farahani, Int. J. Theor. Phys. 53 \textbf{11} 3683 (2014).
\bibitem{12} J. D. Bekenstein, Phys. Rev. D \textbf{7} 2333 (1973).
\bibitem{13} A.F. Heckler. Phys. Rev. D, \textbf{55} 480 (1997).
\bibitem{14} S. Das, P. Majumdar, and R.K. Bhaduri, Class. Quantum Grav.  \textbf{19} 2355(2002).
\bibitem{NPB} B. Pourhassan, Mir Faizal, Nuclear Physics B \textbf{913} 834 (2016).
\bibitem{15} R.K. Kaul, P. Majumdar, Phys. Rev. Lett. \textbf{84} 5255 (2000).
\bibitem{16} P. Majumdar,  Pramana \textbf{55} 511 (2000).
\bibitem{17} J.L. Cardy, Nucl. Phys.  \textbf{B270} 186 (1986).
\bibitem{18} R.K. Kaul, P. Majumdar, Phys. Lett.  \textbf{B439} 267  (1998).
\bibitem{19} S. Carlip, Class. Quantum Gravity \textbf{17} 4175 (2000).
\bibitem{20} T.R. Govindarajan, R.K. Kaul, V. Suneeta, Class. Quantum Gravity \textbf{18} 2877 (2001).
\bibitem{21} R.B. Mann, S.N. Solodukhin, Nucl. Phys.  \textbf{B523} 293 (1998).
\bibitem{22} S.N. Solodukhin, Phys. Rev. D \textbf{57} 2410 (1998).
\bibitem{23} M. Faizal, M.M. Khalil, Int. J. Mod. Phys. A \textbf{30} 1550144 (2015).
\bibitem{24} A.F. Ali, J. High Energy Phys.  \textbf{09} 067 (2012).
\bibitem{25} A. Sen,  J. High Energy Phys. \textbf{04} 156 (2013).
\bibitem{26} B. Pourhassan, M. Faizal,  Europhys. Lett. \textbf{111} 40006 (2015).
\bibitem{27} J. Sadeghi, B. Pourhassan, and F. Rahimi, Canadian Journal of Physics \textbf{92} 1638 (2014).
\bibitem{28} S. Das, R.K. Kaul, P. Majumdar, Phys. Rev. D  \textbf{63} 044019 (2001).
\bibitem{29} M. Faizal, B. Pourhassan,  Phys. Lett. B  \textbf{751} 487 (2015).
\bibitem{30} B. Pourhassan, M. Faizal, Phys. Lett. B  \textbf{755} 444 (2016).
\bibitem{h1} S. S. More, Class. Quant. Grav. \textbf{22} 4129 (2005).
\bibitem{h2} B. Pourhassan, K. Kokabi, Can. J. Phys. \textbf{96} 262 (2018).
\bibitem{h3} B. Pourhassan, K. Kokabi, S. Rangyan, Gen. Relativ. Gravit. \textbf{49} 144 (2017).
\bibitem{h4} B. Pourhassan, K. Kokabi, Int. J. Theor. Phys. \textbf{57} 780 (2018).
\bibitem{h5} B. Pourhassan, K. Kokabi, Z. Sabery,  Annals of Physics \textbf{399} 181 (2018).
\bibitem{h6} S. K. Modak, Phys. Lett. B \textbf{671} 167 (2009).
\bibitem{31} J. Sadeghi, B. Pourhassan,  M. Rostami, Phys. Rev. D  \textbf{94} 064006 (2016).
\bibitem{32} H. Saadat, A. Pourdarvish,  Int. J. Theor. Phys.  \textbf{53} 3014 (2014).
\bibitem{33} R. Banerjee, S.K. Modak, S. Samanta,  Eur. Phys. J. C \textbf{70} 317 (2010).
\bibitem{34} D.V. Fursaev,  Phys. Rev. D \textbf{51} 5352 (1995).
\bibitem{35} M. Cavaglia, S. Das, Class. Quantum Gravity \textbf{ 21 } 4511 (2004).
\bibitem{36} K. Falls, D.F. Litim,  Phys. Rev. D \textbf{ 89 } 084002 (2014).
\bibitem{37} S.W. Hawking,  Commun. Math. Phys. \textbf{ 55 } 133 (1977).
\bibitem{38} M. Zhang,  Nuclear Physics B \textbf{ 935 }170  (2018).
\bibitem{PV} S. Upadhyay, B. Pourhassan, H. Farahani, Phys. Rev. D  \textbf{95} 106014 (2017).
\bibitem{39} D. Kubiznak and R. B. Mann, J. High Energy Phys.  \textbf{ 07 } (2012) 033.
\bibitem{40} S. Dutta, A. Jain, and R. Soni, J. High Energy Phys.  \textbf{ 12 } (2013) 060.
\bibitem{41} C. Niu, Y. Tian, and X. Wu, Phys. Rev. D  \textbf{ 85 } 024017(2012).
\bibitem{42} Y. S. Myung, Phys. Lett. B \textbf{663} 111 (2008).
\bibitem{42-1} S. Upadhyay, B. Pourhassan, PTEP [arXiv:1711.04254]
\bibitem{q1} B. Pourhassan, M. Faizal, S. Capozziello, Annals of Physics \textbf{377} 108 (2017).
\bibitem{q2} B. Pourhassan, M. Faizal, Z. Zaz, and  A. Bhat, Physics Letters B \textbf{773} 325 (2017).
\bibitem{q3} B. Pourhassan, S. Upadhyay, H. Saadat, H. Farahani, Nuclear Physics B \textbf{928} 415 (2018).
\bibitem{44} L. R. Abramo, L. Brenig, E. Gunzig, and A. Saa, Mod. Phys. Lett. A, \textbf{18} 1043 (2003).
\bibitem{45} A. Aceña, A. Anabalon, and D. Astefanesei, Phys. Rev. D, \textbf{87} 124033 (2013).
\bibitem{46} J. Aparicio, D. Grumiller, E. Lopez, I. Papadimitriou, and S. Stricker. J. High Energy Phys.  \textbf{2013 } 128 (2013).
\bibitem{47} O. J. Dias, G. T. Horowitz, and J. E. Santos, J. High Energy Phys.  \textbf{2011 }  115 (2011).
\bibitem{48} Y. Degura, K. Sakamoto, and K. Shiraishi, Grav. Cosmol. \textbf{7} 153 (2001).
\bibitem{49} M. Bañados and S. Theisen, Phys. Rev. D \textbf{72} 064019 (2005).
\bibitem{50} T. Kolyvaris, G. Koutsoumbas, E. Papantonopoulos, and G. Siopsis, Class. Quant. Grav. \textbf{29} 205011 (2012).
\bibitem{51} Y. Brihaye and B. Hartmann, J. High Energy Phys.  \textbf{2012} 50 (2012).
\bibitem{52} A. Anabalón and H. Maeda, Phys. Rev. D \textbf{81} 041501 (2010).
\bibitem{53} U. Nucamendi and M. Salgado, Phys.  Rev.  D \textbf{ 68}, 044026 (2003).
\bibitem{54} F. Correa, C. Martinez, and R. Troncoso, J. High Energy Phys. \textbf{2011} 34 (2011).
\bibitem{55} F. Correa, C. Martinez, and R. Troncoso, J. High Energy Phys.  \textbf{2012} 1 (2012).
\bibitem{BTZ1} M. Ba\~{n}ados, C. Teitelboim, J. Zanelli, Phys. Rev. Lett. \textbf{69} 1849 (1992).
\bibitem{BTZ2} B. Pourhassan, M. Faizal, S. A. Ketabi, Int. J. Mod. Phys. D  \textbf{27} 1850118 (2018).
\bibitem{BTZ3} S. Chougule, S. Dey, B. Pourhassan, M. Faizal, Eur. Phys. J. C \textbf{78} 685  (2018).
\bibitem{EPJC} B. Pourhassan, M. Faizal, Eur. Phys. J. C \textbf{77} 96 (2017).
\bibitem{57} M. S. Ma, R. Zhao, Phys. Lett. B \textbf{751}, 278 (2015).
\bibitem{58} H. B. Callen, John Wiley and Sons, New York,\textbf{NY} USA (1985).
\bibitem{59} B. Pourhassan, M. Faizal, and U. Debnath, Eur. Phys. J. C \textbf{76} 145 (2016).
\bibitem{60} B. Pourhassan, M. Faizal, S. Upadhyay, L. Al Asfar, Eur. Phys. J. C \textbf{77} 555 (2017).
\end{thebibliography}
\end{document}